\documentclass[12pt]{article}

\usepackage{amsmath, amssymb}
\usepackage[margin=0.75in]{geometry}
\usepackage{algorithm, algorithmic}
\usepackage{graphicx}
\usepackage{enumerate}
\usepackage{graphicx}
\newcommand{\indep}{\rotatebox[origin=c]{90}{$\models$}}
\usepackage{multirow}
\usepackage{tikz}
\usetikzlibrary{bayesnet}
\usepackage{subfig}

\DeclareGraphicsExtensions{.pdf,.png,.jpg}

\allowdisplaybreaks
\usepackage[round]{natbib}
\usepackage{siunitx}

\newenvironment{packed_enum}{
\begin{enumerate}
 \setlength{\itemsep}{0pt}
  \setlength{\parskip}{0pt}
  \setlength{\parsep}{0pt}
}{\end{enumerate}}

\begin{document}
\title{A Primer on Causality in Data Science}
\author{Hachem Saddiki and Laura B. Balzer}
\date{January 10, 2019}

\maketitle

\begin{abstract}
Many questions in Data Science are fundamentally causal in that our objective is  to learn the effect of some exposure, randomized or not, on an outcome interest. Even studies that are seemingly non-causal, such as those with the goal of prediction or prevalence estimation, have causal elements, including differential censoring or measurement. As a result, we, as Data Scientists, need to consider the underlying causal mechanisms that gave rise to the data, rather than simply the pattern or association observed in those data. 
In this work, we review the ``Causal Roadmap'' of \cite{Petersen2014roadmap} to provide an introduction to some key concepts in causal inference. Similar to other causal frameworks, the steps of the Roadmap include clearly stating the scientific question, defining of the causal model,  translating the scientific question into a causal parameter, assessing the assumptions needed to express the causal parameter as a statistical estimand, implementation of statistical estimators including parametric and semi-parametric methods, and interpretation of our findings. We believe that using such a framework in Data Science will help to ensure that our statistical analyses are guided by the scientific question driving our research, while avoiding over-interpreting our results. We focus on the effect of an exposure occurring at a single time point and highlight the use of targeted maximum likelihood estimation (TMLE) with Super Learner. 

\end{abstract}

\noindent \textbf{Key words:} Causal inference; Directed acyclic graphs (DAGs);
Observational studies;
Structural causal models;
Targeted learning;
{argeted maximum likelihood estimation (TMLE)

\section{Introduction}

Recently, \cite{Hernan2018b} classified Data Science into three tasks: description, prediction, and causal inference. The first two fall firmly in the realm of statistical inference in that they are purely data-driven tasks, while the last requires something more than the observed data alone \citep{Pearl2016}. Consider, for example, the target population of HIV-infected women of child-bearing age (15-49 years old) in East Africa. After obtaining measurements on sample of women from this population, we could provide some basic descriptive statistics on demographic and clinical variables, such as age, education, use of antiretroviral therapy, pregnancy, and  viral suppression, defined as plasma HIV RNA <500 copies/mL. Likewise, we could use these variables to build a predictor of viral suppression. This predictor could rely on parametric logistic regression or more advanced machine learning algorithms, such as Super Learner \citep{SuperLearner, Petersen2015SL}. 

Now consider the potential impact of pregnancy on clinical outcomes in this population. While optimizing virologic outcomes is essential to preventing mother-to-child-transmission of HIV, the prenatal period could plausibly disrupt or enhance HIV care for a pregnant woman \citep{UNAIDSGap2014}. We can then ask, \textit{"what is the effect of pregnancy on HIV RNA viral suppression among HIV-positive women of child-bearing age in East Africa?"}. While the exposure of pregnancy is not a traditional treatment as commonly considered in a randomized trial, this question is still causal in that we are asking about the outcomes of patients under two different conditions and to answer this question, we must go beyond the observed data set. 

In particular, causal inference requires an \emph{a-priori} specified set of, often untestable, assumptions about the data generating mechanism. Once we posit a causal model, often encoded in the language of causal graphs, we can express our scientific question in terms of a causal quantity. 
Under explicit assumptions, we can then translate that causal quantity into a statistical estimand, a function of the observed data distribution. This translation, called \textit{identifiability}, is not guaranteed, as it depends on the underlying scientific question, the structure of the causal model, and the observed data. Lack of identifiability, however, provides us guidance on further data collection efforts and the additional assumptions needed for such translation. Altogether we obtain a statistical estimand that as closely as possible matches the underlying scientific question
and thereby ensures that our objective is driving the statistical analysis, as opposed to letting the statistical analysis determine the question asked and answered \citep{Petersen2014roadmap, Hernan2008}. Once the estimand has been specified, we return to realm of statistics and the purely data-driven exercises of point estimation, hypothesis testing, and creating $95\%$ confidence intervals. Interpretation of the resulting values, however, requires us again to consider our  causal assumptions.

In this primer, we review the Causal Roadmap of \cite{Petersen2014roadmap} to (1) specify the scientific question; (2) build an accurate causal model of our knowledge; (3) define  the target causal quantity; (4) link the observed data to the causal model; (5) assess identifiability; (6) estimate the resulting statistical parameter; and (7) appropriately interpret the results. This Roadmap borrows the general logic from Descartes's Scientific Method \citep{Descartes1637} and shares a common flow of other causal frameworks \citep{Neyman1923, Rubin1974, Holland1986, Robins1986, Rubin1990, Spirtes93, Pearl2000, Little2000, Dawid2000, Heckman2007, Robins2009, MarkBook, Richardson2013, Hernan2016}. In particular, all approaches demand a clear statement of the research objective, including the target population and interventions of interest \citep{Hernan2018, Ahern2018}. All approaches also provide guidance for conducting a statistical analysis that best answers the motivating  question. Unlike some of the other frameworks, however, the Roadmap emphasizes the use of non-parametric or semi-parametric statistical methods, such as targeted maximum likelihood estimation (TMLE), to avoid unwarranted parametric assumptions and harness recent advances in machine learning. As a result this framework has sometimes been called the Targeted Learning Roadmap \citep{MarkBook, Tran2016, Kreif2017}.

\section{The Roadmap for Causal Inference}

\subsection{Specify the Scientific Question}
\label{sec:question}

The first step is to specify our scientific question. This helps frame our objective in a more detailed way, while incorporating knowledge 
about the study. 
In particular, we need to specify the target population, the exposure, and the outcome of interest.
As our running example, we ask, what is the effect of becoming pregnant on  HIV RNA viral suppression ($<$500 copies/mL) among HIV-positive women of child-bearing age (15-49 years) in East Africa?

This  question 
provides a clear definition of the study variables and objective of our research. 
It also makes explicit that the study only makes claims about the effect of a specific exposure, outcome, and target population. Any claims outside this  context, such as a different exposure,  outcome, or target population, represent distinct questions and would require going through the Roadmap again from the start. The temporal cues present in the research question are of particular importance. They represent the putative cause, here pregnancy, and effect of interest, here viral suppression.
The temporal cues, together with background knowledge, are frequently used as a basis for specifying the causal model, our next step.
 
\subsection{Specify the Causal Model}
\label{sec:causalmodel}

One of the appealing features of causal modeling, and perhaps the reason behind its success, is the rich and flexible language  for encoding mechanisms underlying a data generating process. Here, we focus on \cite{Pearl2000}'s structural causal models, which unify  causal graphs and structural equations \citep{Pearl1988, Goldberger1972, Duncan1975}. 
Structural causal models formalize our knowledge, however limited, of the study, including the relationships between variables and the role of unmeasured factors. 
%

Let us consider again our running example of the impact of pregnancy on HIV viral suppression among women in East Africa. Let $W_1$ denote the set of baseline demographic covariates, such as age, marital status, and education level, and $W_2$ denote the set of pre-exposure HIV care variables, such as prior use of antiretroviral therapy. The exposure $A$ is a binary variable indicating that the woman is known to be pregnant, and the outcome $Y$ is a binary indicator of currently suppressing HIV viral replication: $<$500 copies per mL.
These  constitute the set of \textit{endogenous} variables, which are denoted $X = \{W_1, W_2,A,Y\}$ and are essential to answering the research question.

Each endogenous variable is associated with a latent  background factor $U_{W_1}, U_{W_2},U_A,$ and $U_Y$, respectively. The set of background factors are called \textit{exogenous} variables and denoted $U=(U_{W_1}, U_{W_2},U_A,U_Y)$. These variables account for all other unobserved sources that might influence each of the endogenous variables and can share common components. In our example, unmeasured background factors $U$ might include socioeconomic status, the date of HIV infection, the date of conception, her partner's HIV status, and her genetic profile.

\paragraph{Causal Graphs:} 
The ``causal story'' of the data can be conveyed using the language of graphs \citep{Pearl2000, Pearl2016}. Graphical models consist of a set of nodes representing the variables, and a set of directed or undirected edges connecting these nodes. 
 Two nodes are \emph{adjacent} if there exists an edge between them, and a \emph{path} between two nodes
$A$ and $B$ is a sequence of adjacent nodes starting from $A$ and ending in $B$. If an edge is \emph{directed} from node $A$ to node $B$, then $A$ is the \emph{parent} of $B$, and $B$ is the \emph{child} of $A$. More generally, for any path starting from node $A$, the set of nodes included in this path are \emph{descendants} of $A$, and $A$ is the \emph{ancestor} of all the nodes included in this set.

Here, we are interested in Directed Acyclic Graphs (DAGs), which are fully directed graphs with no path from a given node to itself. DAGs  provide a mechanism to explicitly encode our causal assumptions about the underlying data generating process. Specifically, a variable $A$ is a \emph{direct cause} of another variable $B$, if $B$ is the child of $A$ in the causal graph. Also, a variable $A$ is a \emph{cause} of another variable $B$, if $B$ is a descendant of $A$ in the causal graph \citep{Pearl2000}.


\begin{figure}[ht]
    \centering
    \subfloat[]{{ \begin{tikzpicture}

  \node[latent]  (a) {$\mathbf{A}$};
  \node[latent, above=of a, xshift=-0.5cm]  (w1) {$\mathbf{W_1}$};
  \node[obs, above=of w1, xshift=+1.75cm]  (u) {$\mathbf{U}$};
  \node[latent, right=of a, xshift=+1cm]  (y) {$\mathbf{Y}$};
  \node[latent, above=of y, xshift=+0.5cm] (w2) {$\mathbf{W_2}$};

  \edge {u}{w1,w2,a,y};
  \edge {w1}{w2,a,y};
  \edge {w2}{a,y};
  \edge {a}{y};

\end{tikzpicture} }}%
    \qquad
    \subfloat[]{{ \includegraphics[width=0.4\linewidth]{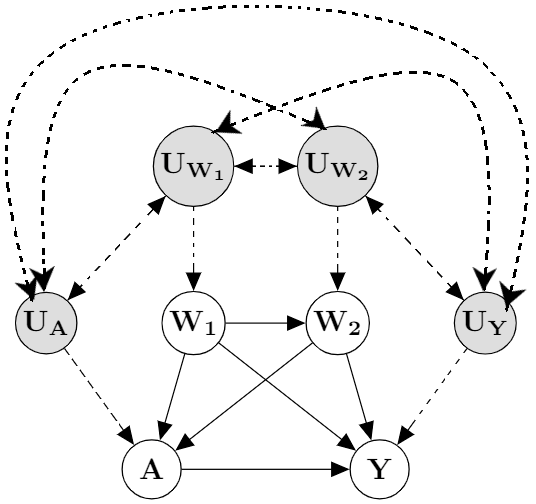} }}%
    	\caption{Encoding the underlying causal mechanisms with graphical models. Shaded nodes represent exogenous variables, and unshaded nodes are endogenous variables. Directed edges represent a direct cause between a pair of variables. Double-headed dashed arrows represent potential correlation between the exogenous factors (i.e., unmeasured common causes of the endogenous variables).
In (a) we give a directed acyclic graph (DAG) with a single node $U$ representing all the common unmeasured sources. In (b)  
we provide an alternative representation to make explicit the relationships between the unmeasured background factors $U=\{U_{W_1}, U_{W_2}, U_A,U_Y\}$ and each endogenous variable. }
            \label{fig:dag1}%
\end{figure}
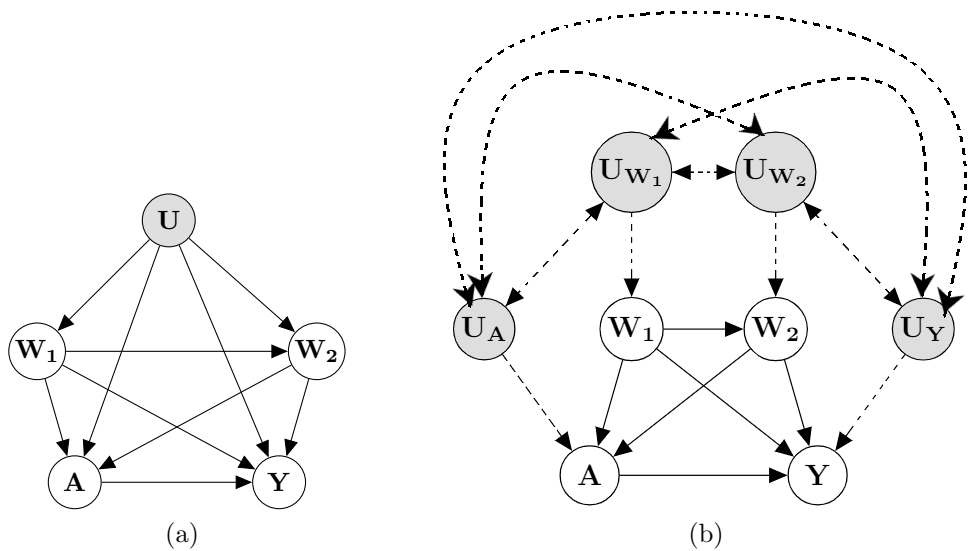

To illustrate, Figure~\ref{fig:dag1}(a) provides a DAG corresponding to our running example. From this graph, we can make the following statements:
\begin{packed_enum}
\item Baseline demographics $W_1$ may affect a woman's pre-exposure HIV care $W_2$, her pregnancy status $A$, and her HIV viral suppression status  $Y$.
\item Prior care $W_2$ may affect  her pregnancy status  $A$, and her HIV viral suppression status  $Y$.
\item Being pregnant $A$ may  affect her HIV viral suppression status $Y$.
\item Unmeasured factors $U=(U_{W_1}, U_{W_2},U_A,U_Y)$ may affect a woman's baseline characteristics, her prior care, her fertility, and her suppression outcome. 
\end{packed_enum}
In Figure~\ref{fig:dag1}(a), a single node $U$ represents all the common, unmeasured factors that could impact the pre-exposure covariates, the exposure, and the outcome. In an alternative representation in Figure~\ref{fig:dag1}(b), we have explicitly shown each exogenous variable ($U_{W_1}, U_{W_2},U_A,U_Y$) as a separate node and as parent to its corresponding endogenous variable $(W_1,W_2,A,Y)$, respectively. In the latter, dashed double-headed arrows denote correlation between the exogenous factors. 

Both representations make explicit that there could be unmeasured common causes of the covariates $W=(W_1,W_2)$ and the exposure $A$, the exposure $A$ and the outcome $Y$, and the covariates $W$ and the outcome $Y$. In other words, there is measured and unmeasured confounding present in this study.   
Altogether, we have avoided many unsubstantiated assumptions about the causal relationships between the variables. This causal model is, thus, non-parametric beyond the assumed time-ordering between variables. 

Causal graphs 
can be extended to accommodate more complicated data structures. Suppose, for example, plasma HIV RNA viral levels are missing for some women in our population of interest. We could modify our causal model to account for incomplete measurement \citep{Robins2000, Robins1994, Scharfstein1999, Daniel2012, Mohan2013, Balzer2017CascadeMethods}. Specifically, we  redefine the exposure node for pregnancy as $A_1$ and introduce a new intervention node $A_2$ defined as indicator that her viral load  is measured. The resulting causal graph is represented in Figure~\ref{fig:dag2}. 
We refer the readers to \cite{Mohan2013} for detailed discussion of formulating a causal model for the missingness mechanism 
and to \cite{PetersenCascade2017} for a real world application handling missingness on both HIV status and viral loads \citep{Balzer2017CascadeMethods}. For the remainder of the primer, we assume, for simplicity, there are no missing data and Figure~\ref{fig:dag1} holds. As discussed in the Appendix, other extensions can also be made to account for common complexities, such as longitudinal data and effect mediation. 

\begin{figure}[ht] 
  \begin{center}
    \begin{tikzpicture}

  \node[latent]  (a1) {$\mathbf{A1}$};
  \node[latent, above=of a1, xshift=-0.5cm]  (w1) {$\mathbf{W_1}$};
\node[latent, right=of w1, xshift=+0.1cm, yshift=-0.5cm]  (w2) {$\mathbf{W_2}$};
  \node[obs, above=of w1, xshift=+1.75cm]  (u) {$\mathbf{U}$};
  \node[latent, right=of a1, xshift=+1cm]  (a2) {$\mathbf{A2}$};
  \node[latent, above=of a2, xshift=+0.5cm, yshift=0.5] (y) {$\mathbf{Y}$};

  \edge {u}{w1, w2,a1,a2,y};
  \edge {w1}{a1,a2,y,w2};
  \edge {w2}{a1,a2,y}
  \edge {a1}{a2,y};
  \edge {a2}{y};

\end{tikzpicture}
  \end{center}
  \caption{Causal graph extending the running example to account for missingness on the outcome. Along the baseline demographic $W_1$, clinical covariates $W_2$, and suppression outcome $Y$, we now have two intervention nodes $A_1$ for pregnancy status and $A_2$ for measurement of plasma HIV RNA level.}
  \label{fig:dag2}
\end{figure}
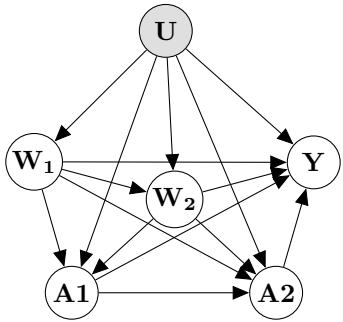


In the subsequent steps, we discuss how altering the causal graph, particularly by removing edges, is equivalent to making additional assumptions about the data generating process. Before doing so, however, we present the causal model in its structural form.

\paragraph{Non-Parametric Structural Equations:}
Structural causal models also encode information about the data generating process with a set of non-parametric equations. Like the causal graph, these equations describe how ``nature'' would deterministically generate the variables in our study  \citep{Pearl2000,Pearl2016}. 
Use of the equations can be preferable in longitudinal settings when  causal graphs can become  unwieldily.

Formally, we define a structural causal model, denoted $\mathcal{M}^*$, by 
the set of exogenous variables $U$, the set of endogenous variables $X$, and a set of functions $\mathcal{F}$ that deterministically assign a value to each variable in $X$, given as input the values of other variables in $X$ and $U$. These non-parametric structural equations allow us to expand our definition of causal assumptions \citep{Pearl2000,Pearl2016}.
Variable $A$ is considered to be a \textit{direct cause} of  variable $B$, if $A$ appears in the function assigning a value to $B$. Variable $A$ is also a \textit{cause} of variable $B$, if $A$ is direct cause of $B$  or any causes   of $B$.

In our HIV viral suppression example, the corresponding structural equations are
\begin{align}
\label{Eq:SCM}
W_1 &= f_{W_1}(U_{W_1}) \nonumber \\
W_2 &= f_{W_2}(W_1, U_{W_2}) \\
A &= f_A(W_1, W_2,U_A) \nonumber \\
Y &= f_Y(W_1, W_2,A,U_Y) \nonumber
\end{align}
where the set of functions $\mathcal{F}=\{f_{W_1}, f_{W_2},f_A,f_Y\}$ encode the mechanism 
deterministically generating the
value of each endogenous variable. 
The exogenous variables $U=\{U_{W_1}, U_{W_2},U_A,U_Y\}$ have a joint probability distribution $\mathbb{P}_U$ and coupled with the set of structural equations $\mathcal{F}$ give rise to a particular data generating process that is compatible with the causal assumptions implied by 
$\mathcal{M}^*$.

In our example, for a given probability distribution $\mathbb{P}_U$ and set of structural equations $\mathcal{F}$, the structural causal model $\mathcal{M}^*$ describes the following data generating process.
For each woman, 
  \begin{packed_enum}
  	\item \textit{Draw the exogenous variables $U$ from the joint probability distribution $\mathbb{P}_U$}. Intuitively, when we sample a woman from the population, we obtain all the unmeasured variables that could influence her baseline covariates, prior care, pregnancy status, and suppression outcome.
    \item \textit{Generate demographic covariates $W_1$ deterministically using $U_{W_1}$ as input to the function $f_{W_1}$}; the demographic covariates include her age, marital status, education attained, and socioeconomic status.
    \item \textit{Generate past HIV care covariates $W_2$ deterministically using $U_{W_2}$ and the woman's demographic covariates $W_1$ as input to the function $f_{W_2}$}; the measured clinical factors include  history of antiretroviral therapy use and prior HIV suppression status.
    \item \textit{Generate pregnancy status $A$ deterministically using $U_A$, $W_1$, and $W_2$ as inputs to function $f_A$}. Recall $A$ is an indicator equaling 1 if the woman is known to be pregnant and 0 otherwise.
    \item \textit{Generate HIV suppression outcome $Y$ deterministically using $U_Y$, $W_1$, $W_2$, and $A$ as inputs to function $f_Y$}.  Recall $Y$ is an indicator equaling 1 if her HIV RNA viral level is less than 500 copies per mL and 0 otherwise.
  \end{packed_enum}
 
It is important to note that the set of structural equations 
are non-parametric. In other words, the explicit relationship between the system variables, as captured by the set of functions $\mathcal{F}$, are left unspecified. If knowledge is available regarding a relationship of interest, it can be readily incorporated in the structural equations. For instance, in a two-armed randomized trial with equal allocation probability, the function that assigns a value to the exposure variable $A$ can be explicitly encoded as $A = f_A(U_A) = \mathbb{I}(U_A < 0.5)$, where $\mathbb{I}$ is an indicator function and $U_A$ assumed to be drawn from a $Uniform(0,1)$.

\subsection{Define the Target Causal Quantity}
\label{sec:causalparameter}

Once the causal model is specified, we may begin to ask questions of causal nature. The rationale comes from the observation that the structural causal model $\mathcal{M}^*$ is not restricted to the particular setting of our study, but can also describe the same system under changed conditions. The structural equations are \emph{autonomous}, which means that modifying one function does not change another. Therefore, we can make targeted modifications to our causal model to evaluate hypothetical, counterfactual scenarios that would otherwise never be realized, but correspond to our underlying scientific question. 

In our running example, we are interested in the effect of pregnancy on  viral suppression. In the original causal model (Figure~\ref{fig:dag1} and  Equation~\ref{Eq:SCM}), a woman's pregnancy status is determined by her baseline demographics $W_1$, prior care status $W_2$, and unmeasured factors $U_A$, such as contraceptive use. However, our objective is to determine the probability of viral suppression if all women in the target population were pregnant versus if the same women over the same time-frame were not pregnant. The autonomy of the structural equations allows us to modify the way in which the exposure, here pregnancy, is determined. In particular,  we can intervene on the exposure $A$  to deterministically set $A = 1$ in one scenario, and then set $A = 0$ in another, while keeping the other equations constant. 

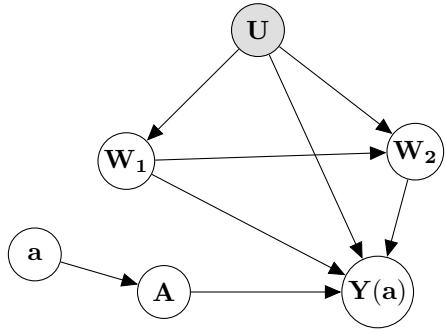
\begin{figure}[ht] 
  \begin{center}
    \begin{tikzpicture}

  \node[latent]  (A) {$\mathbf{A}$};
  \node[latent, above=of A, xshift=-0.5cm]  (w1) {$\mathbf{W_1}$};
  \node[latent, left=of A, yshift=+0.5cm] (a) {$\mathbf{a}$};
  \node[obs, above=of w1, xshift=+1.75cm]  (u) {$\mathbf{U}$};
  \node[latent, right=of A, xshift=+1cm]  (y) {$\mathbf{Y(a)}$};
  \node[latent, above=of y, xshift=+0.5cm] (w2) {$\mathbf{W_2}$};

  \edge {u}{w1,w2,y};
  \edge {w1}{w2,y};
  \edge {w2}{y};
  \edge {A}{y};
  \edge {a}{A};

\end{tikzpicture}
  \end{center}
  \caption{Causal graph after intervention on the exposure, pregnancy status, to set $A=a$. Since this is done deterministically and independently of other variables in the system, the only node $causing$ a change in $A$ is the intervention node $a \in \{0,1\}$. }
  \label{fig:mut_dag1}
\end{figure}

The post-intervention causal graph is given in Figure~\ref{fig:mut_dag1} and the structural equations become
\begin{align*}
W_1 = f_{W_1}(U_{W_1})  &\quad \quad \quad W_1 = f_{W_1}(U_{W_1}) \\
W_2 = f_{W_2}(W_1, U_{W_2})  &\quad \quad \quad W_2 = f_{W_2}(W_1, U_{W_2}) \\
A = 1 &\quad \quad \quad A = 0 \\ 
Y(1) = f_Y(W_1, W_2, 1,U_Y) &\quad \quad \quad Y(0) = f_Y(W_1, W_2,0,U_Y)
\end{align*}
These interventions generate counterfactual outcomes $Y(a)$ for $a\in \{0,1\}$, whose distribution is denoted  $\mathbb{P}^*$. These causal quantities are indicators that a  participant would have suppressed viral replication, if possibly contrary to fact, her pregnancy status were $A=a$.

In this case, it is both physically impossible and unethical to design a randomized trial for pregnancy. In other words, we cannot directly intervene on a woman's pregnancy status. Likewise in Figure~\ref{fig:dag2},  enforcing measurement of the outcome, which translates into setting $A_2=1$, is impossible.  While neither intervention is plausible, we believe counterfactuals provide a language to express many questions in Data Science in a mathematically tractable way. Nonetheless, we note that there has been an lively debate about defining and interpreting analyses of variables on which one cannot directly intervene \citep{Pearl1995, Hernan2005, vanderLaan2005resp, Petersen2014roadmap, Hernan2016}.


Given the counterfactual outcomes and their distribution $\mathbb{P}^*$, we can  express our scientific question as a mathematical quantity. One common choice is the Average Treatment Effect (ATE):
\begin{equation}
\Psi^*(\mathbb{P}^*) := \mathbb{E}^*[Y(1)-Y(0)],
\end{equation}
where the expectation is taken with respect to $\mathbb{P}^*$. Since the causal model $\mathcal{M}^*$ provides the set of possible probability distributions for the exogenous and endogenous factors $(U,X)$ and thus the counterfactual outcomes $(Y(1),Y(0))$, $\Psi^*$ is a mapping from $\mathcal{M}^*$ to the real numbers. 
The target causal parameter 
$\Psi^*(\mathbb{P}^*)$ represents the difference in the expected counterfactual outcome if all units in the target population were exposed and the expected counterfactual outcome if the same units were not exposed.
For the running example, $\Psi^*(\mathbb{P}^*)$ can be interpreted as the difference in the counterfactual probability of viral suppression if all women in the target population were pregnant versus if all women were not. 

Before discussing how these causal quantities can be identified from the observed data distribution, we emphasize that for simplicity we have focused on  a binary intervention, occurring deterministically at a single time point. Scientific questions corresponding to categorical, continuous, stochastic, and longitudinal exposures are also encompassed in this framework, but beyond the scope of this primer and are briefly discussed in the Appendix. We also note that other summaries, such as relative measures, the sample average effect, or marginal structural models,  may better capture the researcher's scientific question. 

\subsection{Link the Observed Data to the Causal Model}
\label{Sec:link}

Thus far, we have defined our scientific question,  specified a structural causal model $\mathcal{M}^*$ to represent our knowledge of the data generating process,  intervened on that causal model to generate counterfactual outcomes, and used these counterfactuals to express our scientific question as a causal quantity.
The next step is to provide an explicit link between the observed data and the specified structural causal model.

Returning to our running example, suppose we have a simple random sample of $N$ women from our target population. On each woman, we measure her baseline demographics $W_1$, prior HIV care $W_2$, pregnancy status $A$, and  suppression outcome $Y$. These measurements constitute our observed data for each woman in our sample: $O=\{W_1, W_2,A,Y\}$. Therefore, we have $N$ independent, identically distributed copies of $O$, which are drawn from some 
probability distribution $\mathbb{P}$.
Other sampling schemes, such as case-control, are accommodated by this framework, but are beyond the scope of this primer.

If we believe that our causal model accurately describes the data generating process, we can assume that the observed data are generated by sampling repeatedly from a distribution compatible with the structural causal model. 
In other words, the structural causal model $\mathcal{M}^*$ provides a description of the study under existing conditions (i.e., the real world) and under specific intervention (i.e., the counterfactual world). As a result, the observed outcome $Y$ equals the counterfactual outcome $Y(a)$ when the observed exposure $A$ equals the exposure of interest, $a$; this is commonly called the \textit{consistency assumption}.

In our example, all the endogenous variables are observed: $X=O$; therefore, we can write
\begin{equation}
\mathbb{P}(O=o) = \sum_{u} \mathbb{P}^*(X=x|U=u)\mathbb{P}^*(U=u),
\end{equation}
where an integral replaces the summation for continuous variables. This, however, might not always be the case. Suppose, for example, we only measured demographics, pregnancy status, and viral suppression, but failed to measure variables related to prior HIV care. Then the observed data would be $O=(W_1,A,Y)$ and are a subset of all the endogenous variables $X$. In either case, we see that the structural causal model $\mathcal{M}^*$, defined as the collection of all possible joint distributions of the exogenous and endogenous variables $(U,X)$, implies the statistical model $\mathcal{M}$, defined as the collection of all possible joint distributions for the observed data $O$. 
The structural causal model $\mathcal{M}^*$ rarely implies restrictions on the resulting statistical model $\mathcal{M}$, which is thereby often non-parametric. An important exception is a completely randomized trial, where  the unmeasured factors determining the treatment assignment $U_A$ are independent of the others and results in a semi-parametric statistical model. The \emph{D-separation} criteria of \cite{Pearl2000} can be used to evaluate what statistical assumptions, if any, are implied by the causal model.
The true observed data distribution 
$\mathbb{P}$
is an element of the statistical model  $\mathcal{M}$.

\subsection{Assessing Identifiability}
\label{sec:ident}

In the previous section, we established a bridge between our structural causal model $\mathcal{M}^*$ and our statistical model $\mathcal{M}$. However, we have not yet discussed the conditions under which causal assumptions and observed data can be combined to answer causal questions. Structural causal models provide one way to assess the assumptions needed to express our target causal quantity as a statistical estimand, which is a well-defined function of the observed data distribution $\mathbb{P}$. 

Recall in Section~\ref{sec:causalparameter} that we defined our target causal parameter as the average treatment effect
$\Psi^* (\mathbb{P}^*)= \mathbb{E}^*[Y(1) - Y(0)]$:
the difference in the expected viral suppression status if all women were pregnant versus if none were. If given a causal model and its link to the observed data, the target causal parameter can be expressed as a function of the observed data distribution $\mathbb{P}$,
then the causal parameter is called $identifiable$. If not, we can still explicitly state and evaluate the assumptions needed to render the target causal parameter identifiable from the observed data distribution.

One of the main tools for assessing identifiability of causal quantities is a set of criteria based on causal graphs. In general, these criteria provide 
a systematic approach to identify an appropriate adjustment set.  
Here, we focus on identifiability for the effect of a single intervention at one time, sometimes called ``point-treatment effects''. For these problems, we first present the back-door criterion and the front-door criterion. For a detailed presentation of graphical methods for assessing identifiability in causal graphs, 
the reader is  referred to \cite{Pearl2000, Pearl2016}.

Formally, we say that a path is \emph{blocked} if at least one variable in that path is conditioned on, and we define a \emph{back-door path} from a given node $A$ as any path that contains an arrow into node $A$. Then, given any pair of variables ($A,B$), where $A$ occurs before $B$ in a directed acyclic graph, a set of variables $C$ is said to satisfy the \emph{back-door criterion} with respect to ($A,B$) if (1) the descendants of $A$ do not include any node in $C$, and (2) $C$ blocks every back-door path from $A$ to $B$ . The rationale behind this criterion is that, for $C$ to be the appropriate adjustment set that isolates the causal effect of $A$ on $B$, we must block all spurious paths between $A$ and $B$, and leave directed paths from $A$ to $B$ unblocked. This criterion does not, however, cover all possible graph structures. 

Alternatively, a set of variables $C$ satisfies the \emph{front-door criterion} with respect to a pair of variables ($A,B$) if (1) all directed paths from $A$ to $B$ are blocked by $C$, (2) all paths from $A$ to $C$ are blocked, and (3) all paths from $C$ to $B$ containing an arrow into $C$ are blocked by $A$. We note that the front-door criterion is more involved than its back-door counterpart, in the sense that it requires more stringent conditions to hold for a given adjustment set to satisfy identifiability. In practice, it is often the case that the back-door criterion is enough to identify the needed adjustment set, 
especially in point-treatment settings. When the back-door criterion holds,  the observed association between the exposure and outcome can be attributed to the causal effect of interest, as opposed to spurious sources of correlation. 
%

In our running example, the set of baseline covariates $W=(W_1, W_2)$ will satisfy the back-door criterion with respect to the effect of pregnancy $A$ on HIV viral suppression $Y$, if the following two conditions hold:
\begin{packed_enum}
\item No node in $W$ is a descendant of $A$.
\item All back-door paths from $A$ to $Y$ are blocked by $W$.
\end{packed_enum}
Looking at the posited causal graph from Figure \ref{fig:dag1}(a), we see that the first condition holds, but the second  is violated. There exists a back-door path  from $A$ to $Y$ through the unmeasured background factors $U$. Intuitively, the unmeasured common causes of pregnancy and HIV viral suppression obstruct our isolation of the causal effect of interest and thus ``confound'' our analyses. Therefore, our target causal quantity is not identifiable in the original causal model $\mathcal{M}^*$. 

Nonetheless, we can explicitly state and consider the plausibility of the causal assumptions needed for identifiability. 
In particular, the following independence assumptions are sufficient to satisfy the back-door criterion and thus identify the causal effect in this point-treatment setting.
\begin{packed_enum}
\item There must not be any unmeasured common causes of the exposure and the outcome: $U_A \indep U_Y$ $and$, 
	\begin{packed_enum}
    \item There must not be any unmeasured common causes of the exposure  and the baseline covariates: $U_A \indep U_{W_1}$ and $U_A \indep U_{W_2}$
    \item[$or$]
    \item There must not be any unmeasured common causes of the baseline covariates and the outcome: $U_{W_1} \indep U_Y$ and $U_{W_2} \indep U_Y$. 
    \end{packed_enum}
\end{packed_enum}
These criteria are reflected in the causal graphs shown in Figure~\ref{fig:ident_dag}. In the running example, assumption 1.a states that there are no unmeasured common causes of pregnancy status and demographic or clinical factors, while 1.b assumes that there are no unmeasured common causes of viral suppression and demographic or clinical factors.

The independence assumptions in 1.a 
hold by design in a stratified, randomized trial, where the unmeasured factors determining the exposure assignment are independent of all other unmeasured factors. As a result, these independence assumptions (1.a and/or 1.b) are sometimes called the \emph{randomization assumption} and equivalently expressed as $Y(a) \indep A \mid W$. These assumptions are also referred to as ``unconfoundedness'', ``selection on observables'', and  ``conditional exchangeability'' \citep{Robins1986}.

\begin{figure}[ht]%
    \centering
    \subfloat[]{{ \begin{tikzpicture}

  \node[latent]  (A) {$\mathbf{A}$};
  \node[latent, above=of A, xshift=-0.5cm]  (w1) {$\mathbf{W_1}$};
  \node[obs, left=of A, yshift=+0.5cm] (ua) {$\mathbf{U_A}$};
  \node[obs, above=of w1, xshift=+1.75cm]  (u) {$\mathbf{U}$};
  \node[latent, right=of A, xshift=+1cm]  (y) {$\mathbf{Y}$};
  \node[latent, above=of y, xshift=+0.5cm] (w2) {$\mathbf{W_2}$};

  \edge {u}{w1,w2,y};
  \edge {w1}{A,w2,y};
  \edge {w2}{A,y};
  \edge {A}{y};
  \edge {ua}{A};

\end{tikzpicture} }}%
    \qquad
    \subfloat[]{{ \begin{tikzpicture}

  \node[latent]  (A) {$\mathbf{A}$};
  \node[latent, above=of A, xshift=-0.5cm]  (w1) {$\mathbf{W_1}$};
  \node[obs, above=of w1, xshift=+1.75cm]  (u) {$\mathbf{U}$};
  \node[latent, right=of A, xshift=+1cm]  (y) {$\mathbf{Y}$};
  \node[obs, right=of y, yshift=+0.5cm] (uy) {$\mathbf{U_Y}$};
  \node[latent, above=of y, xshift=+0.5cm] (w2) {$\mathbf{W_2}$};

  \edge {u}{w1,w2,A};
  \edge {w1}{A,w2,y};
  \edge {w2}{A,y};
  \edge {A}{y};
  \edge {uy}{y};

\end{tikzpicture} }}%
    	\caption{Causal graphs corresponding the identifiability assumptions (1.a) and (1.b), respectively. Here, we have explicitly shown that the unmeasured factors contributing to the exposure $U_A$ in (a) and the outcome $U_Y$ in (b) are independent of the others.}
            \label{fig:ident_dag}%
\end{figure}
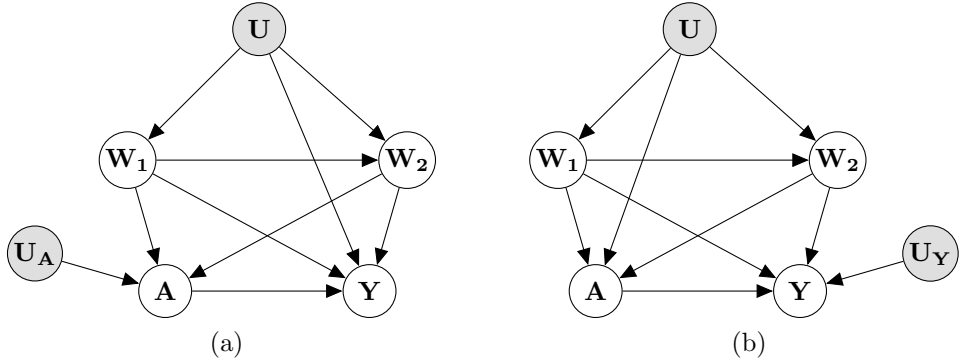

With these assumptions, we can express the distribution of counterfactual outcomes in terms of the distribution of the observed data:
\begin{align*}
\mathbb{P}^*(Y(a)) &= \sum_{w} \mathbb{P}^*(Y(a) | W=w) \mathbb{P}^*(W=w) \\
&= \sum_{w} \mathbb{P}^*(Y(a) | A=a, W=w) \mathbb{P}^*(W=w) \\
&= \sum_{w} \mathbb{P}(Y | A=a, W=w) \mathbb{P}(W=w) 
\end{align*}
where $W=(W_1, W_2)$ denotes the pre-exposure covariates, including both demographic and clinical factors, and where the summation generalizes to an integral for continuous covariates here and in all subsequent expressions.
The first equality is by the law of iterated expectations. The second equality holds by the randomization assumption, and the final by the established link between the causal and statistical model (Section~\ref{Sec:link}).

Under these assumptions, we can express 
the average treatment effect 
$\Psi^*(\mathbb{P}^*)= \mathbb{E}^*[Y(1) - Y(0)]$, 
as a statistical estimand, often called the \emph{G-computation identifiability result} \citep{Robins1986}:
\begin{equation}
\label{stat-estimand}
\Psi(\mathbb{P}) := \sum_{w} \big[ \mathbb{E}(Y|A=1,W=w) - \mathbb{E}(Y|A=0,W=w) \big] \mathbb{P}(W=w)
\end{equation}
Thus, our statistical target is the difference in the expected outcome, given the exposure and covariates, and the expected outcome, given no exposure and covariates, averaged  with respect to the distribution of the baseline covariates $W$. 
In our example, $\Psi(\mathbb{P})$ is the difference in the probability of viral suppression, between pregnant and non-pregnant women with the same values of the covariates, standardized with respect to the covariate distribution in the population.

The same quantity can be expressed in inverse probability weighting form:
\begin{equation}\label{ipw-estimand}
\Psi(\mathbb{P}) := \mathbb{E}\left[ \left( \frac{\mathbb{I}(A=1)}{\mathbb{P}(A=1\mid W)} - \frac{\mathbb{I}(A=0)}{\mathbb{P}(A=0\mid W)}\right) Y \right] \\
\end{equation}
The latter representation highlights an additional data support condition, known as \emph{positivity}:
\begin{equation*}
min_{a\in \mathcal{A}}\ \mathbb{P}(A=a|W=w) > 0, \ \text{for all $w$ such that } \mathbb{P}(W=w)>0.
\end{equation*}
Each exposure level of interest must occur with a positive probability within the strata of the discrete-valued adjustment set $W$. This assumption is also called ``overlap'' and ``the experimental treatment assignment assumption''.
 We refer the reader to \cite{Petersen2012} for a discussion of this assumption and approaches when it is theoretically or practically violated. For continuous covariates, it is not straightforward to evaluate the positivity assumption. One commonly used approach is to transform these continuous variables into categories or quantiles, and assess positivity violations as with categorical variables \citep{Cole2008, Messer2010}. 

Overall, the identifiability step is essential to specifying the needed adjustment set, and thereby statistical estimand to link our causal effect of interest to some function of the observed data distribution. Above, we focused on a simple point-treatment setting with measured and unmeasured confounding, but without mediation, biased sampling, or missing data. In more realistic settings, there many other sources of association between our exposure and outcome, including selection bias, direct and indirect effects, and the  common statistical paradoxes of Berkson's bias and Simpson's Paradox \citep{Hernan2004, HernandezDiaz2006}. Furthermore, in the setting of longitudinal exposures with time-dependent confounding, the needed adjustment set may not be intuitive and the short-comings of traditional approaches become more pronounced \citep{Robins1986,
Robins2000,  
Robins2009, Pearl2016}.
Indeed, methods to distinguish between correlation and causation are crucial in the era of ``Big Data'', where the number of variables is growing with increasing volume, variety, and velocity  \citep{Rose2012problems, Marcus2014problems, Balzer2016BigData}.

Nonetheless, it is important to note that specifying a causal model (Section~\ref{sec:causalmodel}) does not guarantee the identification of a causal effect. Causal frameworks do, however, provide insight into the limitations and full extent of the questions that can be answered given the data at hand. They further facilitate the discussion of modifications to the study design, the measurement additional variables, and sensitivity analyses \citep{Robins1999sensitivity, Imai2010, Vanderweele2011sensitivity, Diaz2012Sensitivity}.

In fact, even if the causal effect is not identifiable (e.g., Figure~\ref{fig:dag1}), the Causal Roadmap still provides us with a statistical estimand (e.g., Equation~\ref{stat-estimand}) that comes as close as possible to the causal effect of interest given the limitations in the observed dataset. In the next sections, we discuss estimation of this statistical parameter and use identifiability results, or lack there of, to inform the strength of our interpretations.

\subsection{Estimate the Target Statistical Parameters}

Once the statistical model and estimand have been defined, the Causal Roadmap returns to traditional statistical inference to estimate functions of a given observed data distribution. Here, we focus on estimators based on the G-computation identifiability result $\Psi(\mathbb{P})$.
Popular methods for estimation and inference for 
$\Psi(\mathbb{P})$,
which would equal the average treatment effect if the identifiability assumptions held, include parametric G-computation, Inverse Probability Weighting (IPW), 
and Targeted Maximum Likelihood Estimation (TMLE) \citep{Robins1986, Horvitz1952, Rosenbaum83, vanderLaan2006, MarkBook}. Below we briefly outline the implementation of each estimator and refer the reader to \cite{PetersenIntroCI} for worked \texttt{R} code for each algorithm. We emphasize that while each algorithm is targeting a causally motivated statistical estimand, these algorithms are not directly estimating causal effects, and therefore it is a misnomer to call them ``causal estimators''. 


\paragraph{Parametric G-computation} is an algorithm that simply estimates the quantities needed to calculate the statistical estimand defined in Equation~\eqref{stat-estimand} and then substitutes those quantities into the G-computation formula \citep{Robins1986, Taubman2009, Young2011, Westreich2012, Zhang2018}. As a result, this algorithm is  sometimes called the \emph{simple substitution estimator} and  is implemented with the following steps.

\begin{packed_enum}
\item Regress the outcome on the exposure and covariate adjustment set to estimate the conditional expectation $\mathbb{E}(Y|A,W)$.
\item Based on the estimates from the Step 1, generate the predicted outcomes for each individual in the sample while deterministically setting the value of the exposure to the levels of interest, but keeping the covariates the same:
\begin{equation*}
\hat{\mathbb{E}}(Y_i | A_i=1, W_i) \text{ and } \hat{\mathbb{E}}(Y_i | A_i=0, W_i) 
\text{ for all observations} \ i=1, ..., N.
\end{equation*}
For a binary outcome this step is corresponds to generating the predicted probabilities $\hat{\mathbb{P}}(Y_i=1 | A_i=a, W_i)$  for exposure levels $a \in\{0,1\}$.
\item 
Obtain a point estimate by taking a sample average of the difference in the predicted outcomes from Step 2:
\begin{equation*}
\hat{\Psi}_{Gcomp}(\hat{\mathbb{P}}) = \frac{1}{N} \sum_{i=1}^N \big[\hat{\mathbb{E}}(Y_i | A_i=1, W_i) -\hat{\mathbb{E}}(Y_i | A_i=0, W_i) \big]
\end{equation*}
where $\hat{\mathbb{P}}$ denotes the empirical distribution, which is the non-parametric maximum likelihood estimator of the covariate  distribution $\mathbb{P}(W)$. The sample proportion is a simple non-parametric estimator of $\mathbb{P}(W=w)$: $\frac{1}{N} \sum_i \mathbb{I}(W_i=w)$.
\end{packed_enum}

\paragraph{Inverse Probability Weighting (IPW)} is an estimator  based on an alternative form of the G-computation identifiability result defined in Equation~\eqref{ipw-estimand} \citep{Horvitz1952, Rosenbaum83, Robins2000, Bodnar2004, Cole2008}. In this form, the  statistical estimand is a function of the conditional probability of being exposed, given the adjustment covariates $\mathbb{P}(A=1|W)$, which is often called the \emph{propensity score} \citep{Rosenbaum83}. IPW controls for confounding by up-weighting rare exposure-covariate subgroups, which have a small propensity score, and down-weighting more common subgroups, which have a larger propensity score. The IPW estimator is implemented with the following steps.

\begin{packed_enum}
\item Regress the exposure on the covariate adjustment set to estimate the propensity score $\mathbb{P}(A=1|W)$.
\item Based on the estimates from Step 1, predict each individual's probability of receiving her observed exposure, given the adjustment covariates:
\begin{equation*}
\hat{\mathbb{P}}(A_i | W_i) 
\text{ for all observations} \ i=1, ..., N.
\end{equation*}
\item Obtain a point estimate by taking the empirical mean of the outcome weighted by the inverse of the conditional exposure probabilities:
\begin{equation*}
\hat{\Psi}_{IPW}(\hat{\mathbb{P}}) = \frac{1}{N} \sum_{i=1}^N \left(\frac{\mathbb{I}(A_i=1)}{\hat{\mathbb{P}}(A_i=1 | W_i)} -\frac{\mathbb{I}(A_i=0)}{\hat{\mathbb{P}}(A_i=0 | W_i)} \right) Y_i.
\end{equation*}
Thus, individuals who are exposed receive weight as one over the estimated propensity score $\hat{\mathbb{P}}(A_i=1 | W_i)$, while individuals who are not exposed receive weight as negative one over the estimated probability of not being exposed, given the covariates $\hat{\mathbb{P}}(A_i=0 | W_i)$. 
\end{packed_enum}

The performance of the parametric G-computation depends on consistent estimation of the conditional expectation of the outcome, given the exposure and covariates $\mathbb{E}(Y|A,W)$, and the performance of IPW relies on consistent estimation of the propensity score $\mathbb{P}(A=1|W)$. Traditionally, both estimators have relied on parametric regression models to estimate these quantities. If sufficient background knowledge is available to support using such a regression, it should have already been encoded in the causal model, yielding parametric structural equations  in Section~\ref{sec:causalmodel}, and can be incorporated during estimation.

However, in most real-world studies with a large number of covariates and potentially complicated relationships, we usually do not have the knowledge support using such parametric regressions. More often, our statistical model $\mathcal{M}$ for the set of possible distributions of the observed data is non-parametric or semi-parametric (Section~\ref{Sec:link}). Furthermore, we want to avoid introducing new and unsubstantiated assumptions during  estimation. Reliance on poorly specified parametric regressions can result in biased point estimates and misleading inference (e.g., \cite{Benkeser2017, LuqueF2018}). At the same time, non-parametric methods, such as stratification, will break down due to sparsity. Here, recent advances in machine learning can help us estimate $\mathbb{E}(Y|A,W)$ and $\mathbb{P}(A=1|W)$ without introducing new assumptions.

\paragraph{Data-adaptive estimation} or machine learning techniques can be used to effectively estimate  the \emph{nuisance parameters}, which are the quantities needed to compute our statistical estimand: $\mathbb{E}(Y|A,W)$ and $\mathbb{P}(A=1|W)$. We focus our discussion on \emph{ensemble learning methods}, which ``stack'' or combine several prediction algorithms together and can be implemented as follows \citep{Wolpert92, Breiman96}. 

First, we pre-specify a library of candidate algorithms, such as  generalized linear models, splines, random forests, neural networks, or support vector machines. We also define a measure of performance through an appropriate loss function, such as the mean squared error or the negative log-likelihood. Next, we randomly split the observed data into training and validation sets to assess the performance of each algorithm in the library. We then fit each algorithm using only data from the training set and predict the outcomes for the units in validation set. Each algorithm's performance is quantified by deviations, corresponding to the loss function, between the actual and predicted outcomes for the units in the validation set. Repeating the process $V$ times amounts to performing \emph{V-fold cross-validation}. We could then select the algorithm with the best performance, corresponding to the smallest cross-validated \emph{risk} estimate. 

This procedure, called Discrete Super Learner \citep{SuperLearner}, effectively sets up a competition between the algorithms specified in the library, and selects the one with the best performance. It naturally follows then that Discrete Super Learner can only perform as well as the best performing algorithm specified in its library. The full Super Learner algorithm improves upon its discrete version by taking a weighted combination of the algorithm-specific predictions to create a new prediction algorithm. We refer the reader to \cite{Chpt3} for further discussion of Super Learner and its properties and to \cite{Naimi2018} for worked examples and \texttt{R} code. The algorithm is  available in the \texttt{SuperLearner} package in \texttt{R} \citep{SLRpkg}.

The goal of Super Learner is to do the best possible job, according to the specified loss function, of predicting the outcome $Y$, given the exposure $A$ and covariates $W$, or predicting the exposure $A$, given the covariates $W$. As a result, Super Learner-based estimators of the nuisance parameters $\mathbb{E}(Y|A,W)$ or $\mathbb{P}(A=1|W)$  have the wrong bias-variance tradeoff for our statistical estimand $\Psi(\mathbb{P})$, which is a single number as opposed to a whole prediction function. TMLE, discussed next, provides one way to integrate data-adaptive algorithms, such as Super Learner, and still obtain the best possible bias-variance tradeoff for the statistical estimand of interest. Indeed, a particular appeal of the Targeted Learning framework is the use of flexible estimation methods to  respect the statistical model, which is often non-parametric, and to minimize the risk of bias due to regression model misspecification. 

\paragraph{Targeted Maximum Likelihood Estimation (TMLE)} provides a general approach to constructing  semi-parametric, efficient, substitution estimators \citep{vanderLaan2006, MarkBook, Petersen2014ltmle}. Here, we provide a very brief overview and refer the reader to \cite{Schuler2017} for a thorough introduction to the algorithm, which is available in the \texttt{tmle}, \texttt{ltmle}, and \texttt{drtmle} packages in \texttt{R} \citep{tmlePackage, ltmlepackage, Benkeser2017}. To implement TMLE for the G-computation identifiability result $\Psi(\mathbb{P})$, given in Equation~\ref{stat-estimand}, we take the following steps.

First, we use Super Learner provide an initial estimator of  the conditional mean outcome, given the exposure and covariates $\mathbb{\hat{E}}^0(Y|A,W)$. Next, we ``target'' this initial estimator  using information from the  propensity score $\hat{\mathbb{P}}(A=1|W)$, also estimated with Super Learner. Informally, this targeting step can be thought of as a second chance to control confounding and serves to reduce statistical bias for the $\Psi(\mathbb{P})$. We denote the updated estimator of the conditional mean outcome as $\mathbb{\hat{E}}^1(Y|A,W)$ and use it to obtain targeted predictions of the outcome  setting the exposures of interest, but keeping the covariates the same:  $\mathbb{\hat{E}}^1(Y_i | A_i=1,W_i)$ and $\mathbb{\hat{E}}^1(Y_i | A_i=0,W_i)$ for all observations $i=1,\ldots, N$.  Finally, we obtain a point estimate by taking the average difference in these targeted predictions. 
\begin{equation*}
    \hat{\Psi}_{TMLE}(\hat{\mathbb{P}}) = \frac{1}{N} \sum_{i=1}^N \big[ \hat{\mathbb{E}}^1 (Y_i|A_i=1,W_i) - \hat{\mathbb{E}}^1 (Y_i|A_i=0,W_i) \big].
\end{equation*}





TMLE's updating step also serves to endow the algorithm with a number of theoretical properties, which often translate into superior performance in finite samples. First, under regularity and empirical process conditions detailed in \cite{MarkBook}, TMLE follows the Central Limit Theorem and thus the normal distribution can be used for  constructing 95\% confidence intervals and hypothesis testing, even if machine learning is used for estimation of the nuisance parameters $\mathbb{E}(Y|A,W)$ or $\mathbb{P}(A=1|W)$. Furthermore, the estimator is \emph{double robust} in that it will be consistent if either  $\mathbb{E}(Y|A,W)$ or $\mathbb{P}(A=1|W)$ is consistently estimated. Collaborative TMLE further improves upon this robustness result \citep{vanderLaan2010ctmle, Gruber2015ctmle}. Finally, if both nuisance parameters are estimated consistently and at fast enough rates, the estimator will be locally efficient and in large samples attain the minimal variance in a semi-parametric statistical model. We refer the reader to \cite{Kennedy2017} for an introduction to semi-parametric, efficiency theory. 

Finally, we note that there is nothing inherent in the TMLE algorithm that demands the use of Super Learner. However, its implementation with machine learning algorithms avoids introducing new unsubstantiated assumptions during estimation and improve our chances for consistent results. Again, relying on misspecified parametric regressions  can induce statistical bias and yield misleading statistical inference.

\subsection{Interpretation of Results}

The final step of the Roadmap is to interpret our results. We have seen that the causal inference framework clearly delineates the assumptions made from domain knowledge (Section~\ref{sec:causalmodel}) from the ones desired for identifiability (Section~\ref{sec:ident}).
In other words, this framework ensures that the assumptions needed to augment the statistical results with a causal interpretation are made explicit. In this regard, \cite{Petersen2014roadmap} argue for a hierarchy of interpretations with ``increasing strength of assumptions''. First, we always have a statistical interpretation as an estimate of the difference in the expected outcome between exposed and unexposed units with the same covariate values, standardized over the covariate distribution in the population. We can also interpret $\hat{\Psi}(\hat{\mathbb{P}})$ as an estimate of the marginal difference in the expected outcome associated with the exposure, after controlling for measured confounding. To interpret our estimates causally, we need the identifiability assumptions (Section~\ref{sec:ident}) to hold in the original causal model (Section~\ref{sec:causalmodel}). If either graphs in Figure~\ref{fig:ident_dag} represented the true causal structure that generated our data and the positivity assumption held, then we could interpret $\hat{\Psi}(\hat{\mathbb{P}})$ as the average treatment effect or for a binary outcome the causal risk difference.
 
Now, recall that the counterfactual outcomes were derived through intervening on the causal model (Section~\ref{sec:causalparameter}). The selected intervention should match our underlying scientific question (Section~\ref{sec:question}) and does not have to correspond to a feasible or realistic intervention. If the identifiability assumptions (Section~\ref{sec:ident}) held and the intervention could be conceivably implemented in the real world, then we could further interpret $\hat{\Psi}(\hat{\mathbb{P}})$ as an estimate of the intervention's impact if it had been implemented in the population of interest. Finally, if the identifiability assumptions were met and the intervention implemented perfectly in a study sample, whose characteristics exactly matched those of our population and who were fully measured, then we could interpret $\hat{\Psi}(\hat{\mathbb{P}})$ as replicating the results of the randomized trial of interest. We note this hierarchy represents a divergence from the Target Trial framework of \cite{Hernan2006b}, who suggest causal inference with observational data can be thought of as ``emulating'' a randomized trial. 

In our running example, the causal model shown in Figure~\ref{fig:dag1} represents our knowledge of the data generating process; there are measured ($W_1,W_2$) as well as unmeasured $U$ common causes of the exposure $A$ and the outcome $Y$. Thus, the lack of identifiability prevents any interpretation as a causal effect or further along the hierarchy. Thus, we can interpret a point estimate of $\Psi(\mathbb{P})$ as the difference in the probability of HIV RNA viral suppression associated with pregnancy after controlling for the measured demographic and clinical confounders.

\section{Conclusion}

The objective of statistical analyses is to make inferences about the data generating process underlying a randomized trial or an observational study. In practice, statistical inference is concerned with purely data-driven tasks, such as prediction, estimation and hypothesis testing. In recent decades, the advent of causal inference has triggered a shift in focus, particularly within the data analysis community, toward a territory that has traditionally evaded statistical reach: the causal mechanism underlying a data generating process. Statistical inference relies on patterns present in the observed data, such as correlation, and  therefore is unable, alone, to answer questions of causal nature \citep{Pearl2010, Pearl2016}.  Nonetheless, questions about cause and effect are of prime importance in all fields including Data Science \citep{Pearl2018, Hernan2018b}.

We have presented an overview of one framework for causal inference.
We emphasized how the Causal Roadmap helps ensure consistency and transparency between the imperfect nature of real world data, and the complexity associated with questions of causal nature. Of course, this work serves only as a primer to causal inference in Data Science, and we have only presented the fundamental concepts and tools in the causal inference arsenal.

Indeed, this framework can be extended to richer and more complicated questions. For instance, our running example for average treatment effect 
only focused on a single exposure at a single time point. However, as demonstrated in \cite{Tran2016, Kreif2017}, the Causal Roadmap can also handle multiple intervention nodes with time-dependent confounding. Other recent avenues of research in causal inference are discussed in the Appendix. 

As a final note, a Data Scientist may debate the usefulness of applying the causal inference machinery to her own research.
We hope to have clarified that if appropriately followed, the Causal Roadmap forces us to think carefully about the goal of our research, the context in which data were collected, and to explicitly define and justify any assumptions. 
It is our belief that conforming to the rigors of this causal inference framework will improve the quality and reproducibility of all scientific endeavors that rely on real data to understand how nature works.

\section*{Appendix}

Here, we briefly highlight some extensions to more advanced settings. For each, we provide a broad definition and a few examples with citations to some relevant works. 
\begin{enumerate}
    \item \textbf{Marginal structural models}  provide a summary  of how the distribution of the counterfactual outcome changes as a function of the exposure and possibly pre-exposure covariates \citep{Robins1999, Robins2000, Bodnar2004, Neugebauer2007, Robins2009, Chpt24, Zheng2016MSM}. 
    Marginal structural models are another way to define our target causal parameter and especially useful when the exposure is continuous or has many levels. \\
    \textit{Examples:} \cite{Robins2000} specified a logistic regression model to summarize the dose-response relation for the cumulative effect of zidovudine (AZT) treatment on the counterfactual risk of having undetectable HIV RNA levels among HIV-positive patients. For a time-to-event outcome, \cite{Cole2012} used a Cox proportional hazard model to summarize the association between treatment initiation and  the counterfactual hazard of incident AIDS or death among persons living with HIV. 

    \item \textbf{Longitudinal exposures}, corresponding to interventions on multiple treatment nodes, allow us to assess the cumulative effect of an exposure or exposures over time \citep{Robins2000,  Bang&Robins05, Robins2009, Chpt24, vanderLaan2012towertmle, Westreich2012, Petersen2014ltmle}. Examining the effects of longitudinal exposures is complicated by  time-dependent confounding, when a covariate is affected by a prior treatment and confounds a future treatment. In these settings, causal frameworks have been especially useful for identifying the appropriate adjustment sets and thereby statistical analysis.\\
 \textit{Examples:} \cite{Schnitzer2014} sought to assess the effect of breastfeeding duration on gastrointestinal infections among new borns, while \cite{Decker2014} investigated the effects of sustained physical activity and diet interventions on adolescent obesity.
  
  \item \textbf{Effect mediation} refers to a general class of causal questions seeking to distinguish an exposure's direct effect on the outcome from its indirect effect through an intermediate variable \citep{Robins1992, Pearl2001, Petersen2006b, vanderLaan2008DE, VanderWeele2009, Imai2010, Zheng2012, Tran2016}.  There are several types of direct and indirect effects. For example, the controlled direct effect refers to the contrast between the expected counterfactual outcomes under two levels of the exposure, but when the mediator is fixed at a constant level. The natural direct effect, also called the pure direct effect, refers to the contrast between the expected counterfactual outcomes under two levels of the exposure, but when the mediator remains at its counterfactual level under the reference value of the exposure. Indirect effects can be defined analogously.  \\
\textit{Examples:} \cite{Naimi2016} examined the  disparity in infant mortality due to race that would remain if all mothers breastfeed prior to hospital discharge. More recently, \cite{Rudolph2018} investigated how the impact of neighborhood disadvantage  on adolescent substance use was mediated by school and peer environment. 

  \item \textbf{Dynamic treatment regimes} are personalized rules for assigning the exposure or treatment as  a  function of an individual's covariate history \citep{Murphy2003, Hernan2006dynamic, vanderLaan2007indvtxt, Kitahata2009, Hernan2009b,  Cain2010, Kreif2017}. They are also called ``adaptive treatment strategies'' and ``individualized treatment rules''. Static interventions, which assign a single level of the exposure to all individuals regardless of their covariate values, can be considered a special case of dynamic interventions.\\
    \textit{Examples:} \cite{Cain2010} and \cite{Young2011} both considered CD4-based thresholds for initiating antiretroviral therapy and their impact on mortality among persons living with HIV.  Recently, \cite{Kreif2017} compared  static and dynamic regimes to understand the optimal timing and level of nutritional support for children in  a pediatric intensive care unit. 
    
\item \textbf{Stochastic interventions} aim to change or shift the  distribution of the exposure \citep{Korb2004, Taubman2009, Cain2010, Diaz2012,  Diaz2013, Rudolph2017}. Stochastic interventions are especially useful when the exposure of interest can not be directly manipulated and can help alleviate violations to the positivity assumption. Deterministic interventions, which assign a given level of the exposure with probability one, can be considered a special case of stochastic interventions. \\
    \textit{Examples}: \cite{Diaz2012} asked what is the impact of a policy encouraging more exercise, according to health and socioeconomic factors, on mortality in a population of older adults? \cite{Danaei2013} examined the impact of various lifestyle interventions, such as eating at least 2 servings of whole grain per day, on the risk of type 2 diabetes in women.

\item \textbf{Clustered data} occur when there is dependence or correlation between individuals within some grouping, such as a clinic, school, neighborhood, or community. Such correlation can arise from shared cluster-level factors, including the exposure, and from social or biological interactions between individuals with a cluster  \citep{Halloran1991, Halloran1995, Oakes2004, Tchetgen2012, vanderLaan2014Networks, Schnitzer2014, Prague2016, Balzer2018Hierarchical, Morozova2018, Buchanan2018}. This dependence must be accounted when specifying the causal model and often demands relaxing the stable unit treatment value assumption,  which prohibits one unit's exposure from impacting another's outcome \citep{Rubin1978}. \\
\textit{Examples:} \cite{Balzer2018Hierarchical} examined the impact of household socioeconomic status, a cluster-level variable, on the risk of failing to test for HIV. Likewise, \cite{Buchanan2018} investigated both the individual and disseminated effects of a network-randomized intervention among people who inject drugs.
    
\item \textbf{Missing data, censoring, and losses to follow up} can all be treated as additional intervention nodes in a given causal framework \citep{Robins2000, Robins1994, Scharfstein1999, Daniel2012, Mohan2013, Balzer2017CascadeMethods}. Thereby, we can treat missing data as a causal inference problem - as opposed to causal inference as a missing data.\\
    \textit{Examples}: When estimating the effect of iron supplementation during pregnancy on anemia at delivery, \cite{Bodnar2004} used inverse probability of censoring weights to adjust for the measured ways in which the women who were censored could differ from those who were not. Likewise, \cite{PetersenCascade2017} estimated the probability of HIV RNA viral suppression over time among a closed cohort of HIV-infected adults, under a hypothetical intervention to prevent censoring and ensure complete viral load measurement. 

    \item \textbf{Transportability}, a subset of generalizability, aims to apply the effect for a given sample to a different population or setting  \citep{Cole2010, Stuart2011, Hernan2011transport, Petersen2011transport, Bareinboim2013, Pearl2015, Lesko2017, Balzer2017general}.  \\
    \textit{Examples:} \cite{Rudolph2017transport} examined whether the reduction in school dropout observed in the Moving to Opportunity trial was consistent between Boston and Los Angeles. Recently, \cite{Hong2018} investigated whether the reductions in cardiovascular risk from rosuvastatin as observed in the JUPITER trial would also have been observed in the UK population who were trial eligible. 
\end{enumerate}

\bibliography{CausalPrimer}

\begin{thebibliography}{125}
\providecommand{\natexlab}[1]{#1}
\providecommand{\url}[1]{\texttt{#1}}
\expandafter\ifx\csname urlstyle\endcsname\relax
  \providecommand{\doi}[1]{doi: #1}\else
  \providecommand{\doi}{doi: \begingroup \urlstyle{rm}\Url}\fi

\bibitem[Ahern(2018)]{Ahern2018}
J.~Ahern.
\newblock Start with the "{C}-word," follow the roadmap for causal inference.
\newblock \emph{American Journal of Public Health}, 108\penalty0 (5):\penalty0
  621, 2018.

\bibitem[Balzer(2017)]{Balzer2017general}
L.~Balzer.
\newblock {``All} generalizations are dangerous, even this one.'' - {Alexandre}
  {Dumas} {[Commentary]}.
\newblock \emph{Epidemiology}, 28\penalty0 (4):\penalty0 562--566, 2017.

\bibitem[Balzer et~al.(2016)Balzer, Petersen, and van~der
  Laan]{Balzer2016BigData}
L.~Balzer, M.~Petersen, and M.~van~der Laan.
\newblock Tutorial for causal inference.
\newblock In P.~Buhlmann, P.~Drineas, M.~Kane, and M.~van~der Laan, editors,
  \emph{Handbook of Big Data}. Chapman \& Hall/CRC, 2016.

\bibitem[Balzer et~al.(2017)Balzer, Schwab, van~der Laan, and
  Petersen]{Balzer2017CascadeMethods}
L.~Balzer, J.~Schwab, M.~van~der Laan, and M.~Petersen.
\newblock Evaluation of progress towards the {UNAIDS} 90-90-90 {HIV} care
  cascade: A description of statistical methods used in an interim analysis of
  the intervention communities in the {SEARCH} study.
\newblock Technical Report 357, University of California at Berkeley, 2017.
\newblock URL \url{http://biostats.bepress.com/ucbbiostat/paper357/}.
\newblock http://biostats.bepress.com/ucbbiostat/paper357/.

\bibitem[Balzer et~al.(2018)Balzer, Zheng, van~der Laan, Petersen, and {the
  SEARCH Collaboration}]{Balzer2018Hierarchical}
L.~Balzer, W.~Zheng, M.~van~der Laan, M.~Petersen, and {the SEARCH
  Collaboration}.
\newblock A new approach to hierarchical data analysis: Targeted maximum
  likelihood estimation for the causal effect of a cluster-level exposure.
\newblock \emph{Stat Meth Med Res}, OnlineFirst, 2018.

\bibitem[Bang and Robins(2005)]{Bang&Robins05}
H.~Bang and J.~Robins.
\newblock Doubly robust estimation in missing data and causal inference models.
\newblock \emph{Biometrics}, 61:\penalty0 962--972, 2005.

\bibitem[Bareinboim and Pearl(2013)]{Bareinboim2013}
E.~Bareinboim and J.~Pearl.
\newblock A general algorithm for deciding transportability of experimental
  results.
\newblock \emph{Journal of Causal Inference}, 1\penalty0 (1):\penalty0
  107--134, 2013.
\newblock \doi{10.1515/jci-2012-0004}.

\bibitem[Benkeser et~al.(2017)Benkeser, Carone, van~der Laan, and
  Gilbert]{Benkeser2017}
D.~Benkeser, M.~Carone, M.~van~der Laan, and P.~Gilbert.
\newblock Doubly robust nonparametric inference on the average treatment
  effect.
\newblock \emph{Biometrika}, 104\penalty0 (4):\penalty0 863--880, 2017.

\bibitem[Bodnar et~al.(2004)Bodnar, Davidian, Siega-Riz, and
  Tsiatis]{Bodnar2004}
L.~Bodnar, M.~Davidian, A.~Siega-Riz, and A.~Tsiatis.
\newblock {Marginal Structural Models for Analyzing Causal Effects of
  Time-dependent Treatments: An Application in Perinatal Epidemiology}.
\newblock \emph{American Journal of Epidemiology}, 159\penalty0 (10):\penalty0
  926--934, 2004.

\bibitem[Breiman(1996)]{Breiman96}
L.~Breiman.
\newblock Stacked regressions.
\newblock \emph{Machine Learning}, 24:\penalty0 49--64, 1996.

\bibitem[Buchanan et~al.(2018)Buchanan, Vermund, Friedman, and
  Spiegelman]{Buchanan2018}
A.~Buchanan, S.~Vermund, S.~Friedman, and D.~Spiegelman.
\newblock Assessing individual and disseminated effects in network-randomized
  studies.
\newblock \emph{Am J Epidemiol}, 187\penalty0 (11):\penalty0 2449--2459, 2018.

\bibitem[Cain et~al.(2010)Cain, Robins, Lanoy, Logan, Costagliola, and
  Hern\'{a}n]{Cain2010}
L.~Cain, J.~Robins, E.~Lanoy, R.~Logan, D.~Costagliola, and M.~Hern\'{a}n.
\newblock When to start treatment? {A} systematic approach to the comparison of
  dynamic regimes using observational data.
\newblock \emph{The International Journal of Biostatistics}, 6\penalty0
  (2):\penalty0 Article 18, 2010.

\bibitem[Cole and Hern\'{a}n(2008)]{Cole2008}
S.~Cole and M.~Hern\'{a}n.
\newblock Constructing inverse probability weights for marginal structural
  models.
\newblock \emph{American Journal of Epidemiology}, 168\penalty0 (6):\penalty0
  656--664, 2008.

\bibitem[Cole and Stuart(2010)]{Cole2010}
S.~Cole and E.~Stuart.
\newblock Generalizing evidence from randomized clinical trials to target
  populations: the {ACTG} 320 {Trial}.
\newblock \emph{American Journal of Epidemiology}, 172\penalty0 (1):\penalty0
  107--115, 2010.
\newblock \doi{10.1093/aje/kwq084}.

\bibitem[Cole et~al.(2012)Cole, Hudgens, Tien, Anastos, Kingsley, Chmiel, and
  Jacobson]{Cole2012}
S.~Cole, M.~Hudgens, P.~Tien, K.~Anastos, L.~Kingsley, J.~Chmiel, and
  L.~Jacobson.
\newblock Marginal structural models for case-cohort study designs to estimate
  the association of antiretroviral therapy initiation with incident {AIDS} or
  death.
\newblock \emph{Am J Epidemiol}, 175\penalty0 (5):\penalty0 381--390, 2012.

\bibitem[Danaei et~al.(2013)Danaei, Pan, Hu, and Hern\'{a}n]{Danaei2013}
G.~Danaei, A.~Pan, F.~Hu, and M.~Hern\'{a}n.
\newblock Hypothetical midlife interventions in women and risk of type 2
  diabetes.
\newblock \emph{Epidemiol}, 24\penalty0 (1):\penalty0 122--128, 2013.

\bibitem[Daniel et~al.(2012)Daniel, Kenward, Cousens, and {De
  Stavola}]{Daniel2012}
R.~Daniel, M.~Kenward, S.~Cousens, and B.~{De Stavola}.
\newblock Using causal diagrams to guide analysis in missing data problems.
\newblock \emph{Stat Meth Med Res}, 21\penalty0 (3):\penalty0 243--256, 2012.

\bibitem[Dawid(2000)]{Dawid2000}
A.~Dawid.
\newblock Causal inference without counterfactuals.
\newblock \emph{Journal of the American Statistical Association}, 95\penalty0
  (450):\penalty0 407--424, 2000.

\bibitem[Decker et~al.(2014)Decker, Hubbard, Crespi, Seto, and
  Wang]{Decker2014}
A.~Decker, A.~Hubbard, C.~Crespi, E.~Seto, and M.~Wang.
\newblock Semiparametric estimation of the impacts of longitudinal
  interventions on adolescent obesity using targeted maximum-likelihood:
  Accessible estimation with the ltmle package.
\newblock \emph{Journal of Causal Inference}, 2\penalty0 (1):\penalty0 95--108,
  2014.

\bibitem[Descartes(1637)]{Descartes1637}
R.~Descartes.
\newblock \emph{Discours de la M{\'e}thode Pour bien conduire sa raison, et
  chercher la v{\'e}rit{\'e} dans les sciences}.
\newblock Leiden, Netherlands, 1637.

\bibitem[D\'{i}az and van~der Laan(2012)]{Diaz2012}
I.~D\'{i}az and M.~van~der Laan.
\newblock Population intervention causal effects based on stochastic
  interventions.
\newblock \emph{Biometrics}, 68\penalty0 (2):\penalty0 541--549, 2012.

\bibitem[D\'{i}az and van~der Laan(2013{\natexlab{a}})]{Diaz2012Sensitivity}
I.~D\'{i}az and M.~van~der Laan.
\newblock Sensitivity analysis for causal inference under unmeasured
  confounding and measurement error problems.
\newblock \emph{Int J Biostat}, 9:\penalty0 149--160, 2013{\natexlab{a}}.

\bibitem[D\'{i}az and van~der Laan(2013{\natexlab{b}})]{Diaz2013}
I.~D\'{i}az and M.~van~der Laan.
\newblock Assessing the causal effect of policies: An example using stochastic
  interventions.
\newblock \emph{Int J Biostat}, 9\penalty0 (2):\penalty0 161--174,
  2013{\natexlab{b}}.

\bibitem[Duncan(1975)]{Duncan1975}
O.~Duncan.
\newblock \emph{Introduction to Structural Equation Models}.
\newblock Academic Press, New York, 1975.

\bibitem[Goldberger(1972)]{Goldberger1972}
A.~Goldberger.
\newblock Structural equation models in the social sciences.
\newblock \emph{Econometrica: Journal of the Econometric Society}, 40:\penalty0
  979--1001, 1972.

\bibitem[Gruber and van~der Laan(2012)]{tmlePackage}
S.~Gruber and M.~van~der Laan.
\newblock {tmle}: An {R} package for targeted maximum likelihood estimation.
\newblock \emph{Journal of Statistical Software}, 51\penalty0 (13):\penalty0
  1--35, 2012.
\newblock \doi{10.18637/jss.v051.i13}.

\bibitem[Gruber and van~der Laan(2015)]{Gruber2015ctmle}
S.~Gruber and M.~van~der Laan.
\newblock Consistent causal effect estimation under dual misspecification and
  implications for confounder selection procedures.
\newblock \emph{Stat Methods Med Res}, 24\penalty0 (6):\penalty0 1003--1008,
  2015.
\newblock PMID: 22368176.

\bibitem[Halloran and Struchiner(1991)]{Halloran1991}
M.~Halloran and C.~Struchiner.
\newblock Study designs for dependent happenings.
\newblock \emph{Epidemiology}, 2:\penalty0 331--338, 1991.

\bibitem[Halloran and Struchiner(1995)]{Halloran1995}
M.~Halloran and C.~Struchiner.
\newblock Causal inference in infectious diseases.
\newblock \emph{Epidemiology}, 6\penalty0 (2):\penalty0 142--151, 1995.

\bibitem[Heckman and Vytlacil(2007)]{Heckman2007}
J.~Heckman and E.~Vytlacil.
\newblock Econometric evaluation of social programs, part {I}: causal models,
  structural models and econometric policy evaluation.
\newblock \emph{Handbook of Econometrics}, pages 4779--4874, 2007.

\bibitem[Hern\'{a}n(2005)]{Hernan2005}
M.~Hern\'{a}n.
\newblock Invited commentary: hypothetical interventions to define causal
  effects--afterthought or prerequisite?
\newblock \emph{Am J Epidemiol}, 162\penalty0 (7):\penalty0 618--620, 2005.

\bibitem[Hern\'{a}n(2018)]{Hernan2018}
M.~Hern\'{a}n.
\newblock The {C}-word: Scientific euphemisms do not improve causal inference
  from observational data.
\newblock \emph{American Journal of Public Health}, 108\penalty0 (5):\penalty0
  616--619, 2018.

\bibitem[Hern\'{a}n and Robins(2006)]{Hernan2006b}
M.~Hern\'{a}n and J.~Robins.
\newblock Estimating causal effects from epidemiological data.
\newblock \emph{J Epidemiol Community Health}, 60\penalty0 (7):\penalty0
  578--586, 2006.

\bibitem[Hern\'{a}n and Robins(2009)]{Hernan2009b}
M.~Hern\'{a}n and J.~Robins.
\newblock {Comment on: Early versus deferred antiretroviral therapy for HIV on
  survival}.
\newblock \emph{New England Journal of Medicine}, 361\penalty0 (8):\penalty0
  823--824, 2009.

\bibitem[Hern\'{a}n and Robins(2016)]{Hernan2016}
M.~Hern\'{a}n and J.~Robins.
\newblock Using big data to emulate a target trial when a randomized trial is
  not available.
\newblock \emph{American Journal of Epidemiology}, 183\penalty0 (8):\penalty0
  758--764, 2016.

\bibitem[Hern\'{a}n and VanderWeele(2011)]{Hernan2011transport}
M.~Hern\'{a}n and T.~VanderWeele.
\newblock Compound treatments and transportability of causal inference.
\newblock \emph{Epidemiology}, 22:\penalty0 368--377, 2011.

\bibitem[Hern\'{a}n et~al.(2004)Hern\'{a}n, Hern\'{a}ndez-D\'{i}az, and
  Robins]{Hernan2004}
M.~Hern\'{a}n, S.~Hern\'{a}ndez-D\'{i}az, and J.~Robins.
\newblock A structural approach to selection bias.
\newblock \emph{Epidemiology}, 15\penalty0 (5):\penalty0 615--625, 2004.

\bibitem[Hern\'{a}n et~al.(2006)Hern\'{a}n, Lanoy, Costagliola, and
  Robins]{Hernan2006dynamic}
M.~Hern\'{a}n, E.~Lanoy, D.~Costagliola, and J.~Robins.
\newblock Comparison of dynamic treatment regimes via inverse probability
  weighting.
\newblock \emph{Basic \& Clinical Pharmacology \& Toxicology}, 98\penalty0
  (3):\penalty0 237--242, 2006.

\bibitem[Hern\'{a}n et~al.(2008)Hern\'{a}n, Alonso, Logan, Grodstein, Michels,
  Willett, Manson, and Robins]{Hernan2008}
M.~Hern\'{a}n, A.~Alonso, R.~Logan, F.~Grodstein, K.~Michels, W.~Willett,
  J.~Manson, and J.~Robins.
\newblock Observational studies analyzed like randomized experiments: an
  application to postmenopausal hormone therapy and coronary heart disease.
\newblock \emph{Epidemiology}, 19:\penalty0 766--779, 2008.

\bibitem[Hern\'{a}n et~al.(2018)Hern\'{a}n, Hsu, and Healy]{Hernan2018b}
M.~Hern\'{a}n, J.~Hsu, and B.~Healy.
\newblock Data science is science's second chance to get causal inference
  right: A classification of data science tasks.
\newblock Technical report, arXiv, 2018.
\newblock URL \url{https://arxiv.org/abs/1804.10846}.
\newblock https://arxiv.org/abs/1804.10846.

\bibitem[{Hern\'{a}ndez-D\'{i}az} et~al.(206){Hern\'{a}ndez-D\'{i}az},
  Schisterman, and Hern\'{a}n]{HernandezDiaz2006}
S.~{Hern\'{a}ndez-D\'{i}az}, E.~Schisterman, and M.~Hern\'{a}n.
\newblock The birth weight ``paradox'' uncovered?
\newblock \emph{Am J Epidemiol}, 164\penalty0 (11):\penalty0 1115--1120, 206.

\bibitem[Holland(1986)]{Holland1986}
P.~Holland.
\newblock Statistics and causal inference.
\newblock \emph{{Journal of the American Statistical Association}}, 81\penalty0
  (396):\penalty0 945--960, 1986.

\bibitem[Hong et~al.(2018)Hong, {Jonsson Funk}, {LoCasale}, Dempster, Cole,
  {Webster-Clark}, Edwards, and Sturmer]{Hong2018}
J.~Hong, M.~{Jonsson Funk}, R.~{LoCasale}, S.~Dempster, S.~Cole,
  M.~{Webster-Clark}, J.~Edwards, and T.~Sturmer.
\newblock Generalizing randomized clinical trial results: Implementation and
  challenges related to missing data in the target population.
\newblock \emph{Am J Epidemiol}, 184\penalty0 (4):\penalty0 817--827z, 2018.

\bibitem[Horvitz and Thompson(1952)]{Horvitz1952}
D.~Horvitz and D.~Thompson.
\newblock A generalization of sampling without replacement from a finite
  universe.
\newblock \emph{{Journal of the American Statistical Association}},
  47:\penalty0 663--685, 1952.
\newblock \doi{10.2307/2280784}.

\bibitem[Imai et~al.(2010)Imai, Keele, and Yamamoto]{Imai2010}
K.~Imai, L.~Keele, and T.~Yamamoto.
\newblock Identification, inference, and sensitivity analysis for causal
  mediation effects.
\newblock \emph{Statistical Science}, 25:\penalty0 51--71, 2010.

\bibitem[{Joint United Nations Programme on HIV/AIDS
  (UNAIDS)}(2014)]{UNAIDSGap2014}
{Joint United Nations Programme on HIV/AIDS (UNAIDS)}.
\newblock The gap report.
\newblock Geneva, Switzerland, 2014.

\bibitem[Kennedy(2017)]{Kennedy2017}
E.~Kennedy.
\newblock Semiparametric theory.
\newblock Technical report, arXiv, 2017.
\newblock URL \url{https://arxiv.org/abs/1709.06418v1}.
\newblock https://arxiv.org/abs/1709.06418v1.

\bibitem[Kitahata et~al.(2009)Kitahata, Gange, Abraham, Merriman, Saag,
  Justice, et~al.]{Kitahata2009}
M.~Kitahata, S.~Gange, A.~Abraham, B.~Merriman, M.~Saag, A.~Justice, et~al.
\newblock {Effect of early versus deferred antiretroviral therapy for HIV on
  survival}.
\newblock \emph{New England Journal of Medicine}, 360\penalty0 (18):\penalty0
  1815--1826, 2009.

\bibitem[Korb et~al.(2004)Korb, Hope, Nicholson, and Axnick]{Korb2004}
K.~Korb, L.~Hope, A.~Nicholson, and K.~Axnick.
\newblock Varieties of causal intervention.
\newblock In C.~Zhang, H.~Guesgen, and W.~Yeap, editors, \emph{PRICAI 2004:
  Trends in Artificial Intelligence, volume 3157 of Lecture Notes in Computer
  Science}, pages 322--331. Springer, Heidelberg, Germany, 2004.

\bibitem[Kreif et~al.(2017)Kreif, Tran, Grieve, {De Stavola}, Tasker, and
  Petersen]{Kreif2017}
N.~Kreif, L.~Tran, R.~Grieve, B.~{De Stavola}, R.~Tasker, and M.~Petersen.
\newblock Estimating the comparative effectiveness of feeding interventions in
  the pediatric intensive careunit: A demonstration of longitudinal targeted
  maximum likelihood estimation.
\newblock \emph{American Journal of Epidemiology}, 186\penalty0 (12):\penalty0
  1370--1379, 2017.

\bibitem[Lendle et~al.(2017)Lendle, Schwab, Petersen, and {van der
  Laan}]{ltmlepackage}
S.~Lendle, J.~Schwab, M.~Petersen, and M.~{van der Laan}.
\newblock ltmle: An {R} package implementing targeted minimum loss-based
  estimation for longitudinal data.
\newblock \emph{Journal of Statistical Software}, 81\penalty0 (1):\penalty0
  1--21, 2017.

\bibitem[Lesko et~al.(2017)Lesko, Buchanan, Westreich, Edwards, Hudgens, and
  Cole]{Lesko2017}
C.~Lesko, A.~Buchanan, D.~Westreich, J.~Edwards, M.~Hudgens, and S.~Cole.
\newblock Generalizing study results: a potential outcomes perspective.
\newblock \emph{Epidemiology}, 28\penalty0 (4):\penalty0 553--561, 2017.

\bibitem[Little and Rubin(2000)]{Little2000}
R.~Little and D.~Rubin.
\newblock Causal effects in clinical and epidemiological studies via potential
  outcomes: concepts and analytical approaches.
\newblock \emph{Annual Revue of Public Health}, 21:\penalty0 121--145, 2000.

\bibitem[Luque-Fernandez et~al.(2018)Luque-Fernandez, Belot, Valeri, Cerulli,
  Maringe, and Rachet]{LuqueF2018}
M.~Luque-Fernandez, A.~Belot, L.~Valeri, G.~Cerulli, C.~Maringe, and B.~Rachet.
\newblock Data-adaptive estimation for double-robust methods in
  population-based cancer epidemiology: Risk differences for lung cancer
  mortality by emergency presentation.
\newblock \emph{American Journal of Epidemiology}, 187\penalty0 (4):\penalty0
  871--878, 2018.

\bibitem[Marcus and Davis(2014)]{Marcus2014problems}
G.~Marcus and E.~Davis.
\newblock Eight (no, nine!) problems with big data.
\newblock \emph{The New York Times}, 2014.
\newblock URL
  \url{http://www.nytimes.com/2014/04/07/opinion/eight-no-nine-problems-with-big-data.html}.

\bibitem[Messer et~al.(2010)Messer, Oakes, and Mason]{Messer2010}
L.~Messer, J.~Oakes, and S.~Mason.
\newblock Effects of socioeconomic and racial residential segregation on
  preterm birth: a cautionary tale of structural confounding.
\newblock \emph{American Journal of Epidemiology}, 171:\penalty0 664--673,
  2010.

\bibitem[Mohan et~al.(2013)Mohan, Pearl, and Tian]{Mohan2013}
K.~Mohan, J.~Pearl, and J.~Tian.
\newblock Graphical models for inference with missing data.
\newblock In C.~J.~C. Burges, L.~Bottou, M.~Welling, Z.~Ghahramani, and K.~Q.
  Weinberger, editors, \emph{Advances in Neural Information Processing Systems
  26}, pages 1277--1285. Curran Associates, Inc., 2013.
\newblock URL
  \url{http://papers.nips.cc/paper/4899-graphical-models-for-inference-with-missing-data.pdf}.

\bibitem[Morozova et~al.(2018)Morozova, Cohen, and Crawford]{Morozova2018}
O.~Morozova, T.~Cohen, and F.~Crawford.
\newblock Risk ratios for contagious outcomes.
\newblock \emph{J. R. Soc. Interface}, 15\penalty0 (20170696), 2018.

\bibitem[Murphy(2003)]{Murphy2003}
S.~Murphy.
\newblock Optimal dynamic treatment regimes.
\newblock \emph{J R Stat Soc Ser B}, 65\penalty0 (2):\penalty0 331--355, 2003.

\bibitem[Naimi and Balzer(2018)]{Naimi2018}
A.~Naimi and L.~Balzer.
\newblock Stacked generalization: An introduction to super learning.
\newblock \emph{European Journal of Epidemiology}, pages 459--464, 2018.

\bibitem[Naimi et~al.(2016)Naimi, Schnitzer, Moodie, and Bodnar]{Naimi2016}
A.~Naimi, M.~Schnitzer, E.~Moodie, and L.~Bodnar.
\newblock Mediation analysis for health disparities research.
\newblock \emph{Am J Epidemiol2016}, 184\penalty0 (4):\penalty0 315--324, 2016.

\bibitem[Neugebauer and van~der Laan(2007)]{Neugebauer2007}
R.~Neugebauer and M.~J. van~der Laan.
\newblock Nonparametric causal effects based on marginal structural models.
\newblock \emph{Journal of Statistical Planning and Inference}, 137\penalty0
  (2):\penalty0 419--434, 2007.

\bibitem[Neyman(1923)]{Neyman1923}
J.~Neyman.
\newblock Sur les applications de la theorie des probabilites aux experiences
  agricoles: Essai des principes ({In Polish}). {E}nglish translation by {D.M.}
  {D}abrowska and {T.P.} {S}peed (1990).
\newblock \emph{Statistical Science}, 5:\penalty0 465--480, 1923.

\bibitem[Oakes(2004)]{Oakes2004}
J.~Oakes.
\newblock The (mis)estimation of neighborhood effects: causal inference for a
  practicable social epidemiology (with discussion).
\newblock \emph{Soc Sci Med}, 58\penalty0 (10):\penalty0 1929--1952, 2004.
\newblock PMID: 15020009.

\bibitem[Pearl(1988)]{Pearl1988}
J.~Pearl.
\newblock \emph{Probabilistic Reasoning in Intelligent Systems}.
\newblock Morgan Kaufmann, San Mateo, CA, 1988.

\bibitem[Pearl(1995)]{Pearl1995}
J.~Pearl.
\newblock Causal diagrams for empirical research.
\newblock \emph{Biometrika}, 82:\penalty0 669--710, 1995.
\newblock \doi{10.1093/biomet/82.4.669}.

\bibitem[Pearl(2000)]{Pearl2000}
J.~Pearl.
\newblock \emph{{Causality: Models, Reasoning and Inference}}.
\newblock Cambridge University Press, New York, 2000.
\newblock Second ed., 2009.

\bibitem[Pearl(2001)]{Pearl2001}
J.~Pearl.
\newblock Direct and indirect effects.
\newblock In \emph{Proceedings of the Seventeenth Conference on Uncertainty in
  Artificial Intelligence}, pages 411--420, San Francisco, 2001. Morgan
  Kaufmann.

\bibitem[Pearl(2010)]{Pearl2010}
J.~Pearl.
\newblock An introduction to causal inference.
\newblock \emph{The International Journal of Biostatistics}, 6\penalty0
  (2):\penalty0 Article 7, 2010.

\bibitem[Pearl(2015)]{Pearl2015}
J.~Pearl.
\newblock Generalizing experimental findings.
\newblock \emph{Journal of Causal Inference}, 3\penalty0 (2):\penalty0
  259--266, 2015.

\bibitem[Pearl(2018)]{Pearl2018}
J.~Pearl.
\newblock The seven tools of causal inference with reflections on machine
  learning.
\newblock Technical Report R-481, UCLA, 2018.

\bibitem[Pearl et~al.(2016)Pearl, Glymour, and Jewell]{Pearl2016}
J.~Pearl, M.~Glymour, and N.~Jewell.
\newblock \emph{Causal inference in statistics: a primer}.
\newblock John Wiley and Sons Ltd, Chichester, West Sussex, UK, 2016.

\bibitem[Petersen(2011)]{Petersen2011transport}
M.~Petersen.
\newblock Compound treatments, transportability, and the structural causal
  model: the power and simplicity of causal graphs.
\newblock \emph{Epidemiology}, 22:\penalty0 378--381, 2011.

\bibitem[Petersen and Balzer(2014)]{PetersenIntroCI}
M.~Petersen and L.~Balzer.
\newblock Introduction to causal inference. {UC} {B}erkeley.
\newblock www.ucbbiostat.com/labs, Aug 2014.

\bibitem[Petersen and van~der Laan(2011)]{Chpt24}
M.~Petersen and M.~van~der Laan.
\newblock Case {S}tudy: {L}ongitudinal {HIV} {C}ohort {D}ata.
\newblock In M.~van~der Laan and S.~Rose, editors, \emph{Targeted Learning:
  Causal Inference for Observational and Experimental Data}. Springer, New York
  Dordrecht Heidelberg London, 2011.

\bibitem[Petersen and van~der Laan(2014)]{Petersen2014roadmap}
M.~Petersen and M.~van~der Laan.
\newblock Causal models and learning from data: Integrating causal modeling and
  statistical estimation.
\newblock \emph{Epidemiology}, 25\penalty0 (3):\penalty0 418--426, 2014.

\bibitem[Petersen et~al.(2006)Petersen, Sinisi, and van~der
  Laan]{Petersen2006b}
M.~Petersen, S.~Sinisi, and M.~van~der Laan.
\newblock Estimation of direct causal effects.
\newblock \emph{Epidemiology}, 17\penalty0 (3):\penalty0 276--284, 2006.

\bibitem[Petersen et~al.(2012)Petersen, Porter, Gruber, Wang, and van~der
  Laan]{Petersen2012}
M.~Petersen, K.~Porter, S.~Gruber, Y.~Wang, and M.~van~der Laan.
\newblock Diagnosing and responding to violations in the positivity assumption.
\newblock \emph{Statistical Methods in Medical Research}, 21\penalty0
  (1):\penalty0 31--54, 2012.
\newblock \doi{10.1177/0962280210386207}.

\bibitem[Petersen et~al.(2014)Petersen, Schwab, Gruber, Blaser, Schomaker, and
  van~der Laan]{Petersen2014ltmle}
M.~Petersen, J.~Schwab, S.~Gruber, N.~Blaser, M.~Schomaker, and M.~van~der
  Laan.
\newblock Targeted maximum likelihood estimation for dynamic and static
  longitudinal marginal structural working models.
\newblock \emph{Journal of Causal Inference}, 2\penalty0 (2), 2014.
\newblock \doi{10.1515/jci-2013-0007}.

\bibitem[Petersen et~al.(2015)Petersen, {LeDell}, Schwab, Sarovar, Gross,
  Reynolds, et~al.]{Petersen2015SL}
M.~Petersen, E.~{LeDell}, J.~Schwab, V.~Sarovar, R.~Gross, N.~Reynolds, et~al.
\newblock Super learner analysis of electronic adherence data improves viral
  prediction and may provide strategies for selective {HIV} {RNA} monitoring.
\newblock \emph{J Acquir Immune Defic Syndr}, 69\penalty0 (1):\penalty0
  109--118, 2015.

\bibitem[Petersen et~al.(2017)Petersen, Balzer, Kwarsiima, Sang,
  et~al.]{PetersenCascade2017}
M.~Petersen, L.~Balzer, D.~Kwarsiima, N.~Sang, et~al.
\newblock Association of implementation of a universal testing and treatment
  intervention with {HIV} diagnosis, receipt of antiretroviral therapy, and
  viral suppression among adults in {East} {Africa}.
\newblock \emph{JAMA}, 317\penalty0 (21):\penalty0 2196--2206, 2017.
\newblock \doi{10.1001/jama.2017.5705}.

\bibitem[Polley et~al.(2011)Polley, Rose, and van~der Laan]{Chpt3}
E.~Polley, S.~Rose, and M.~van~der Laan.
\newblock {Super Learner}.
\newblock In M.~van~der Laan and S.~Rose, editors, \emph{Targeted Learning:
  Causal Inference for Observational and Experimental Data}. Springer, New York
  Dordrecht Heidelberg London, 2011.

\bibitem[Polley et~al.(2018)Polley, LeDell, Kennedy, and {van der
  Laan}]{SLRpkg}
E.~Polley, E.~LeDell, C.~Kennedy, and M.~{van der Laan}.
\newblock \emph{SuperLearner: Super Learner Prediction}, 2018.
\newblock URL \url{https://CRAN.R-project.org/package=SuperLearner}.
\newblock R package version 2.0-24.

\bibitem[Prague et~al.(2016)Prague, Wang, Stephens, {E. Tchetgen Tchetgen}, and
  {De Gruttola}]{Prague2016}
M.~Prague, R.~Wang, A.~Stephens, {E. Tchetgen Tchetgen}, and V.~{De Gruttola}.
\newblock Accounting for interactions and complex inter-subject dependency in
  estimating treatment effect in cluster-randomized trials with missing
  outcomes.
\newblock \emph{Biometrics}, 72\penalty0 (4):\penalty0 1066--1077, 2016.
\newblock \doi{10.1111/biom.12519}.

\bibitem[Richardson and Robins(2013)]{Richardson2013}
T.~Richardson and J.~Robins.
\newblock Single world intervention graphs ({SWIG}s): A unification of the
  counterfactual and graphical approaches to causality.
\newblock Working paper number 128, Center for Statistics and the Social
  Sciences University of Washington, 2013.

\bibitem[Robins(1986)]{Robins1986}
J.~Robins.
\newblock A new approach to causal inference in mortality studies with
  sustained exposure periods--application to control of the healthy worker
  survivor effect.
\newblock \emph{Mathematical Modelling}, 7:\penalty0 1393--1512, 1986.
\newblock \doi{10.1016/0270-0255(86)90088-6}.

\bibitem[Robins(1999)]{Robins1999}
J.~Robins.
\newblock {Association, Causation, and Marginal Structural Models}.
\newblock \emph{Synthese}, 121\penalty0 (1-2):\penalty0 151--179, 1999.

\bibitem[Robins and Hern\'{a}n(2009)]{Robins2009}
J.~Robins and M.~Hern\'{a}n.
\newblock {Estimation of the causal effects of time-varying exposures}.
\newblock In G.~Fitzmaurice, M.~Davidian, G.~Verbeke, and G.~Molenberghs,
  editors, \emph{{Longitudinal Data Analysis}}, chapter~23. Chapman \&
  Hall/CRC, Boca Raton, FL, 2009.

\bibitem[Robins and Rotnitzky(1992)]{Robins1992}
J.~Robins and A.~Rotnitzky.
\newblock Recovery of information and adjustment for dependent censoring using
  surrogate markers.
\newblock In N.~Jewell, K.~Dietz, and V.~Farewell, editors, \emph{{AIDS
  Epidemiology - Methodological Issues}}, Boston, 1992. Birkh\"{a}user.

\bibitem[Robins et~al.(1994)Robins, Rotnitzky, and Zhao]{Robins1994}
J.~Robins, A.~Rotnitzky, and L.~Zhao.
\newblock Estimation of regression coefficients when some regressors are not
  always observed.
\newblock \emph{{Journal of the American Statistical Association}},
  89:\penalty0 846--866, 1994.
\newblock \doi{10.2307/2290910}.

\bibitem[Robins et~al.(1999)Robins, Rotnitzky, and
  Scharfstein]{Robins1999sensitivity}
J.~Robins, A.~Rotnitzky, and D.~Scharfstein.
\newblock Sensitivity analysis for selection bias and unmeasured confounding in
  missing data and causal inference models.
\newblock In M.~Halloran and D.~Berry, editors, \emph{Statistical Models in
  Epidemiology: The Environment and Clinical Trials}. Springer, New York, 1999.

\bibitem[Robins et~al.(2000)Robins, Hern\'{a}n, and Brumback]{Robins2000}
J.~Robins, M.~Hern\'{a}n, and B.~Brumback.
\newblock Marginal structural models and causal inference in epidemiology.
\newblock \emph{Epidemiology}, 11\penalty0 (5):\penalty0 550--560, 2000.

\bibitem[Rose(2012)]{Rose2012problems}
S.~Rose.
\newblock Big data and the future.
\newblock \emph{Significance}, 9\penalty0 (4):\penalty0 47--48, 2012.

\bibitem[Rosenbaum and Rubin(1983)]{Rosenbaum83}
P.~Rosenbaum and D.~Rubin.
\newblock The central role of the propensity score in observational studies.
\newblock \emph{Biometrika}, 70:\penalty0 41--55, 1983.

\bibitem[Rubin(1974)]{Rubin1974}
D.~Rubin.
\newblock Estimating causal effects of treatments in randomized and
  nonrandomized studies.
\newblock \emph{Journal of Educational Psychology}, 66\penalty0 (5):\penalty0
  688--701, 1974.
\newblock \doi{10.1037/h0037350}.

\bibitem[Rubin(1978)]{Rubin1978}
D.~B. Rubin.
\newblock Bayesian inference for causal effects: the role of randomization.
\newblock \emph{Ann Stat}, 6:\penalty0 34--58, 1978.

\bibitem[Rubin(1990)]{Rubin1990}
D.~B. Rubin.
\newblock Comment: {N}eyman (1923) and causal inference in experiments and
  observational studies.
\newblock \emph{Statistical Science}, 5\penalty0 (4):\penalty0 472--480, 1990.

\bibitem[Rudolph and {van der Laan}(2017)]{Rudolph2017transport}
K.~Rudolph and M.~{van der Laan}.
\newblock Robust estimation of encouragement-design intervention effects
  transported across sites.
\newblock \emph{J R Stat Soc Ser B}, 79\penalty0 (5):\penalty0 1509--1525,
  2017.

\bibitem[Rudolph et~al.(2017)Rudolph, Sofrygin, Zheng, and {van der
  Laan}]{Rudolph2017}
K.~Rudolph, O.~Sofrygin, W.~Zheng, and M.~{van der Laan}.
\newblock Robust and flexible estimation of stochastic mediation effects: a
  proposed method and example in a randomized trial setting.
\newblock \emph{Epidemiologic Methods}, 7, 07 2017.

\bibitem[Rudolph et~al.(2018)Rudolph, Goin, Paksarian, Crowder, Merikangas, and
  Stuart]{Rudolph2018}
K.~Rudolph, D.~Goin, D.~Paksarian, R.~Crowder, K.~Merikangas, and E.~Stuart.
\newblock Causal mediation analysis with observational data: Considerations and
  illustration examining mechanisms linking neighborhood poverty to adolescent
  substance use.
\newblock \emph{Am J Epidemiol}, Epub Ahead of Print, 2018.

\bibitem[Scharfstein et~al.(1999)Scharfstein, Rotnitzky, and
  Robins]{Scharfstein1999}
D.~Scharfstein, A.~Rotnitzky, and J.~Robins.
\newblock {Adjusting for Nonignorable Drop-Out Using Semiparametric Nonresponse
  Models (with Rejoiner)}.
\newblock \emph{{Journal of the American Statistical Association}}, 94\penalty0
  (448):\penalty0 1096--1120 (1135--1146), 1999.
\newblock \doi{10.2307/2669930}.

\bibitem[Schnitzer et~al.(2014)Schnitzer, van~der Laan, Moodie, and
  Platt]{Schnitzer2014}
M.~Schnitzer, M.~van~der Laan, E.~Moodie, and R.~Platt.
\newblock Effect of breastfeeding on gastrointestinal infection in infants: a
  targeted maximum likelihood approach for clustered longitudinal data.
\newblock \emph{Annals of Applied Statistics}, 8\penalty0 (2):\penalty0
  703--725, 2014.

\bibitem[Schuler and Rose(2017)]{Schuler2017}
M.~Schuler and S.~Rose.
\newblock Targeted maximum likelihood estimation for causal inference in
  observational studies.
\newblock \emph{American Journal of Epidemiology}, 185\penalty0 (1):\penalty0
  65--73, 2017.

\bibitem[Spirtes et~al.(1993)Spirtes, Glymour, and Scheines]{Spirtes93}
P.~Spirtes, C.~Glymour, and R.~Scheines.
\newblock \emph{Causation, Prediction and Search. Number 81 in Lecture Notes in
  Statistics}.
\newblock Springer-Verlag, New York/Berlin, 1993.

\bibitem[Stuart et~al.(2011)Stuart, Cole, Bradshaw, and Leaf]{Stuart2011}
E.~Stuart, S.~Cole, C.~Bradshaw, and P.~Leaf.
\newblock The use of propensity scores to assess the generalizability of
  results from randomized trials.
\newblock \emph{Journal of the Royal Statistical Society: Series A},
  174\penalty0 (Part 2):\penalty0 369--386, 2011.
\newblock \doi{10.1111/j.1467-985X.2010.00673.x}.

\bibitem[Taubman et~al.(2009)Taubman, Robins, Mittleman, and
  Hern\'{a}n]{Taubman2009}
S.~Taubman, J.~Robins, M.~Mittleman, and M.~Hern\'{a}n.
\newblock Intervening on risk factors for coronary heart disease: an
  application of the parametric {G-formula}.
\newblock \emph{International Journal of Epidemiology}, 38\penalty0
  (6):\penalty0 1599--1611, 2009.

\bibitem[{Tchetgen Tchetgen} and VanderWeele(2012)]{Tchetgen2012}
E.~{Tchetgen Tchetgen} and T.~VanderWeele.
\newblock On causal inference in the presence of interference.
\newblock \emph{Stat Meth Med Res}, 21\penalty0 (1):\penalty0 55--75, 2012.

\bibitem[Tran et~al.(2016)Tran, Yiannoutsos, Musick, Wools-Kaloustian, Siika,
  Kimaiyo, {van der Laan}, and Petersen]{Tran2016}
L.~Tran, C.~Yiannoutsos, B.~Musick, K.~Wools-Kaloustian, A.~Siika, S.~Kimaiyo,
  M.~{van der Laan}, and M.~Petersen.
\newblock Evaluating the impact of a {HIV} low-risk express care task-shifting
  program: A case study of the targeted learning roadmap.
\newblock \emph{Epidemiologic Methods}, 5\penalty0 (1):\penalty0 69--91, 2016.

\bibitem[van~der Laan(2014)]{vanderLaan2014Networks}
M.~van~der Laan.
\newblock Causal inference for a population of causally connected units.
\newblock \emph{Journal of Causal Inference}, 0\penalty0 (0):\penalty0 1--62,
  2014.
\newblock \doi{10.1515/jci-2013-0002}.

\bibitem[van~der Laan and Gruber(2010)]{vanderLaan2010ctmle}
M.~van~der Laan and S.~Gruber.
\newblock Collaborative double robust targeted maximum likelihood estimation.
\newblock \emph{The International Journal of Biostatistics}, 6\penalty0 (1),
  2010.
\newblock \doi{10.2202/1557-4679.1181}.

\bibitem[van~der Laan and Gruber(2012)]{vanderLaan2012towertmle}
M.~van~der Laan and S.~Gruber.
\newblock Targeted minimum loss based estimation of causal effects of multiple
  time point interventions.
\newblock \emph{The International Journal of Biostatistics}, 8\penalty0 (1),
  2012.

\bibitem[van~der Laan and Petersen(2007)]{vanderLaan2007indvtxt}
M.~van~der Laan and M.~Petersen.
\newblock Causal effect models for realistic individualized treatment and
  intention to treat rules.
\newblock \emph{The International Journal of Biostatistics}, 3\penalty0
  (1):\penalty0 Article 3, 2007.

\bibitem[van~der Laan and Petersen(2008)]{vanderLaan2008DE}
M.~van~der Laan and M.~Petersen.
\newblock Direct effect models.
\newblock \emph{The International Journal of Biostatistics}, 4\penalty0
  (1):\penalty0 Article 23, 2008.

\bibitem[van~der Laan and Rose(2011)]{MarkBook}
M.~van~der Laan and S.~Rose.
\newblock \emph{Targeted Learning: Causal Inference for Observational and
  Experimental Data}.
\newblock Springer, New York Dordrecht Heidelberg London, 2011.

\bibitem[van~der Laan and Rubin(2006)]{vanderLaan2006}
M.~van~der Laan and D.~Rubin.
\newblock Targeted maximum likelihood learning.
\newblock \emph{The International Journal of Biostatistics}, 2\penalty0
  (1):\penalty0 Article 11, 2006.
\newblock \doi{10.2202/1557-4679.1043}.

\bibitem[{van der Laan} et~al.(2005){van der Laan}, Haight, and
  Tager]{vanderLaan2005resp}
M.~{van der Laan}, T.~Haight, and I.~Tager.
\newblock van der {Laan} et al. respond to "hypothetical interventions to
  define causal effects''.
\newblock \emph{Am J Epidemiol}, 162\penalty0 (7):\penalty0 621--622, 2005.

\bibitem[van~der Laan et~al.(2007)van~der Laan, Polley, and
  Hubbard]{SuperLearner}
M.~van~der Laan, E.~Polley, and A.~Hubbard.
\newblock Super learner.
\newblock \emph{Statistical Applications in Genetics and Molecular Biology},
  6\penalty0 (1):\penalty0 25, 2007.
\newblock \doi{10.2202/1544-6115.1309}.

\bibitem[VanderWeele(2009)]{VanderWeele2009}
T.~VanderWeele.
\newblock Marginal structural models for the estimation of direct and indirect
  effects.
\newblock \emph{Epidemiology}, 20\penalty0 (1):\penalty0 18--26, 2009.

\bibitem[VanderWeele and Arah(2011)]{Vanderweele2011sensitivity}
T.~VanderWeele and O.~Arah.
\newblock Bias formulas for sensitivity analysis of unmeasured confounding for
  general outcomes, treatments, and confounders.
\newblock \emph{Epidemiology}, 22:\penalty0 42--52, 2011.

\bibitem[Westreich et~al.(2012)Westreich, Cole, Young, Palella, Tien, Kingsley,
  Gange, and Hern\'{a}n]{Westreich2012}
D.~Westreich, S.~Cole, J.~Young, F.~Palella, P.~Tien, L.~Kingsley, S.~Gange,
  and M.~Hern\'{a}n.
\newblock The parametric g-formula to estimate the effect of highly active
  antiretroviral therapy on incident aids or death.
\newblock \emph{Statistics in Medicine}, 31\penalty0 (18):\penalty0 2000--2009,
  2012.
\newblock \doi{10.1002/sim.5316}.

\bibitem[Wolpert(1992)]{Wolpert92}
D.~H. Wolpert.
\newblock Stacked generalization.
\newblock \emph{Neural Networks}, 5:\penalty0 241--259, 1992.

\bibitem[Young et~al.(2011)Young, Cain, Robins, O'Reilly, and
  Hern\'{a}n]{Young2011}
J.~Young, L.~Cain, J.~Robins, E.~O'Reilly, and M.~Hern\'{a}n.
\newblock Comparative effectiveness of dynamic treatment regimes: An
  application of the parametric g-formula.
\newblock \emph{Stat Biosci}, 3:\penalty0 119--143, 2011.

\bibitem[Zhang et~al.(2018)Zhang, Young, Thamer, and Hern\'{a}n]{Zhang2018}
Y.~Zhang, J.~Young, M.~Thamer, and M.~Hern\'{a}n.
\newblock Comparing the effectiveness of dynamic treatment strategies using
  electronic health records: An application of the parametric g-formula to
  anemia management strategies.
\newblock \emph{Health Serv Res}, 53\penalty0 (3):\penalty0 1900--1918, 2018.

\bibitem[Zheng and van~der Laan(2012)]{Zheng2012}
W.~Zheng and M.~van~der Laan.
\newblock Targeted maximum likelihood estimation for natural direct effects.
\newblock \emph{The International Journal of Biostatistics}, 8\penalty0
  (1):\penalty0 1--40, 2012.

\bibitem[Zheng et~al.(2016)Zheng, Petersen, and {van der Laan}]{Zheng2016MSM}
W.~Zheng, M.~Petersen, and M.~{van der Laan}.
\newblock Doubly robust and efficient estimation of marginal structural models
  for the hazard function.
\newblock \emph{Int J Biostat}, 12\penalty0 (1):\penalty0 233--252, 2016.

\end{thebibliography}

\end{document}